%% file: aimerson_galform_clusters_accepted.tex
\documentclass[useAMS,usenatbib]{mn2e}

\pdfoutput=1
\usepackage{times}  % to get nice fonts for PDF
\usepackage{url}
\usepackage{graphics}
\usepackage{graphicx}
\usepackage{amssymb}
\usepackage[draft]{hyperref}
\usepackage{aas_macros}
\usepackage[dvipsnames]{xcolor}
\usepackage{lastpage}
\newcommand{\GALFORM}{{\tt GALFORM}}

\newcommand{\Euclid}{\emph{Euclid}}

\newcommand{\Mpc}{{\rm Mpc}}
\newcommand{\Msolyr}{{\rm M_{\odot}yr^{-1}}}
\newcommand{\Msol}{{\rm M_{\odot}}}

\newcommand{\arcseconds}{\prime\prime}
\newcommand{\band}[1]{{\rm #1}}

\input{epsf} \title{The abundance and colours of galaxies in high
  redshift clusters in the cold dark matter cosmology}

\author[Merson {\it et al.}]
{\parbox[h]{\textwidth}{Alexander~I.~Merson$^{1}$\thanks{E-mail:
      alexander.merson@ucl.ac.uk}, Carlton~M.~Baugh$^2$,
    Violeta~Gonzalez-Perez$^2$,\\Filipe~B.~Abdalla$^{1,3}$,
    Claudia~del~P.~Lagos$^{4,5}$, Simona~Mei$^{6,7,8}$}
  \vspace*{3pt}\\
  \noindent$^1$Department of Physics and Astronomy, University College
  London, Gower Street, London, WC1E 6BT\\$^2$Institute for
  Computational Cosmology (ICC), Department of Physics, Durham
  University, South Road, Durham, DH1 3LE\\$^3$Department of Physics
  and Electronics, Rhodes University, PO Box 94, Grahamstown, 6140,
  South Africa\\$^4$European Southern Observatory,
  Karl-Schwarzschild-Strasse 2, D-85748 Garching,
  Germany\\$^5$International Centre for Radio Astronomy Research
  (ICRAR), M468, University of Western Australia, 35 Stirling Hwy,
  Crawley, WA 6009, Australia\\$^6$GEPI,
  Observatoire de Paris, Section de Meudon, 5 Place J. Janssen, 92190
  Meudon Cedex, France\\$^7$Universit\'{e} Paris Denis Diderot, 75205
  Paris Cedex 13, France\\$^8$Infrared Processing and Analysis
  Center, California Institute of Technology, Pasadena, CA 91125,
  USA}\date{}

\pagerange{\pageref{firstpage}--\pageref{lastpage}}
\pubyear{\the\year}

\begin{document}

\maketitle
\title{Abundance and colours of cluster galaxies}
\label{firstpage}

\begin{abstract}
  High redshift galaxy clusters allow us to examine galaxy formation
  in extreme environments. Here we compile data for 15 $z>1$ galaxy
  clusters to test the predictions from a state-of-the-art
  semi-analytical model of galaxy formation. The model gives a good
  match to the slope and zero-point of the cluster red sequence. The
  model is able to match the cluster galaxy luminosity function at
  faint and bright magnitudes, but under-estimates the number of
  galaxies around the break in the cluster luminosity function. We
  find that simply assuming a weaker dust attenuation improves the
  model predictions for the cluster galaxy luminosity function, but
  worsens the predictions for the red sequence at bright
  magnitudes. Examination of the properties of the bright cluster
  galaxies suggests that the default dust attenuation is large due to
  these galaxies having large reservoirs of cold gas as well as small
  radii. We find that matching the luminosity function and colours of
  high redshift cluster galaxies, whilst remaining consistent with
  local observations, poses a challenge for galaxy formation models.
\end{abstract}


%%%%%%%%%%%%%%%%%%%%%%%%%%%%%%%%%%%%%%%%%%%%%%%%%%%%%%%%%%%%%%%%%%%%%%%%%%%%%%%%%%%%%%%%%%%%%%%%%%%
% INTRODUCTION
%%%%%%%%%%%%%%%%%%%%%%%%%%%%%%%%%%%%%%%%%%%%%%%%%%%%%%%%%%%%%%%%%%%%%%%%%%%%%%%%%%%%%%%%%%%%%%%%%%%

\section{Introduction}
\label{sec:intro}

Galaxy clusters are the most massive bound structures found in the
Universe. Not only are clusters excellent proxies for massive dark
matter halos (and therefore a useful cosmological probe), but they are
also unique sites of galaxy evolution \citep[e.g. ][]{Kravtsov12}

Over the past decade a wealth of observational data has been gathered
that points towards the redshift range $z\sim1.5-2.5$ as being a
pivotal epoch in the evolutionary history of the galaxy population,
with star formation, black hole accretion and galaxy mergers reaching
their peak activity before being subsequently suppressed
\citep[e.g.][]{Dickinson03,Hopkins04}. Similarly, observations
indicate that at this epoch galaxy clusters and proto-clusters were in
the process of being transformed from dynamical, merger-driven
over-densities to the more relaxed systems that we see today. As such,
there is mounting evidence that this redshift range marks the
quenching of star formation in massive cluster galaxies and the
build-up of the cluster red sequence \citep[RS,
][]{Bower92b,Bower92a,Lidman08,Brammer11}. However, developing a
physical description of the rapid evolution of the galaxy population
in clusters in this transformation phase represents a challenge for
current models of galaxy formation.

The next generation of cosmological galaxy surveys, such as the
\emph{Dark Energy Survey} \citep[DES,][]{DES05} or the European
Space Agency's \Euclid{} mission \citep{Laureijs11}, are expected to
observe many thousands of high redshift galaxy clusters. However,
identifying galaxy clusters in the huge volumes probed by these
surveys is a difficult task fraught with systematics, especially as
these surveys will be predominantly photometric. The large
uncertainties inherent in photometric redshift estimation make the
identification of cluster members based upon their spatial separation,
as is the case for the \emph{friends-of-friends} \citep{Huchra82} and
\emph{Voroni-Delaney} methods \citep{Marinoni02}, much more
challenging \citep[e.g.][]{Zandivarez14}.

A more favourable approach for photometric datasets is to identify
cluster members based upon whether they lie on the cluster RS in the
colour-magnitude relation (CMR). Under the assumption that early-type
galaxies dominate the cluster galaxy population and that this
population follows a tight relation in colour-magnitude space, then,
when imaged in two photometric bands bracketing the $4000{\rm \AA}$
break, the cluster galaxies will be the brightest, reddest objects
\citep{Stanford98,Gladders00}. The observational efficiency of this
approach has led to it being used extensively in cluster detection
\citep{Gladders05,Wilson06} and adopted in several group and cluster
finding algorithms \citep[e.g.][]{Koester07,Murphy12,Rykoff14}. As
discussed in \citeauthor{Gladders00}, optical and infrared imaging of
local and $z>1$ galaxy clusters indicates the universal presence of a
RS \citep[e.g.][]{Baldry04,Bell04,Brinchmann04}, with many studies
supporting a high formation redshift of the stellar population of
$z_f\gtrsim2$ \citep[e.g.][]{Ellis97, Smail98, Stanford98, Ponman99,
  Lopez-Cruz04, Gladders05, Miller05, Voit05, Mei06a, Mei06b,
  Koester07, Gilbank08, Lidman08, Mei09, Wilson09, Gilbank11, Lin12,
  Mei12}. There is, however, some observational evidence for ongoing
star-formation in clusters at $z\gtrsim1$, \citep[e.g.][]{Hayashi10,
  Brodwin13, Fassbender14, Mei14}.

Although using galaxy colours to identify galaxy clusters is
preferable when dealing with the larger uncertainties inherent in
photometric redshifts \citep[e.g.][]{Abdalla11}, this approach is
sensitive to our understanding of the astrophysical processes
governing the evolution of galaxies in cluster environments and the
build-up of the RS, as well as possible biases introduced by
photometric colour selections. As a result, the RS method must be
tested and calibrated. At low redshift this can be done with
spectroscopic datasets, but at higher redshifts, where the
spectroscopic data is sparse, one must turn to using synthetic `mock'
catalogues based upon the latest galaxy formation models
\citep{Baugh08}.

Due to their rarity, generating a population of galaxy clusters
requires very large volume cosmological N-body simulations, which due
to limitations in computational resources are typically dark matter
only simulations. Several techniques are available for populating the
halos from N-body simulations with galaxies. A common approach is to
use empirical methods, such as the \emph{halo occupation distribution}
\citep[HOD,][]{Berlind02} or \emph{sub-halo abundance matching}
\citep{Vale04}. These methods have the benefit that they are tuned
using observations to ensure that the luminosity function and colour
distribution of the galaxies are correct by construction. At high
redshifts, however, the lack of observational data prohibits the use
of such methods, though some redshift dependent approaches have been
proposed \citep[e.g. ][]{Moster13}.

Instead, \emph{semi-analytic} galaxy formation models provide a more
flexible alternative \citep{Baugh06}. These models use simple
prescriptions to describe the various physical processes governing the
evolution of the baryon content of the halos and aim to predict the
fundamental properties of galaxies, such as their stellar mass and
star formation history, ab initio. Adoption of a stellar population
synthesis model and a choice of initial stellar mass function, allows
one to translate these fundamental properties into directly observable
properties. Although semi-analytic models still require observational
data to constrain their parameters\footnote{Unlike HODs, the
  observational data used to constrain semi-analytic models need not
  be from the particular epoch of interest. Semi-analytic models can
  be constrained using local observations and then be used to make
  high redshift predictions.}, they have been shown to make realistic
predictions for the evolution of the global galaxy population out to
high redshift \citep[e.g.][]{Lacey11}. However, given the extreme
cosmic evolution of galaxy clusters, which account for only a few per
cent of all mass today, making accurate predictions for the properties
of high redshift cluster galaxies remains a challenge as we shall see.

Here we examine the predictions made by a semi-analytical galaxy
formation model for the near-infrared (near-IR) photometry of galaxies
in clusters at redshifts $z>1$. Our decision to examine the statistics
of clusters at this epoch in the near-infrared is motivated by the
aims of the \Euclid{} mission, which will provide a deep survey over
15,000 ${\rm deg}^2$ of the sky to a photometric depth of
$\band{H}\lesssim24$. As such, \Euclid{} is predicted to provide a
uniformly selected sample of approximately 60,000 clusters with a
signal-to-noise greater than 3, with approximately 10,000 of these
lying at $z>1$ \citep{Laureijs11}. Examining the near-IR predictions
for this epoch is therefore extremely timely in the preparation for
\Euclid{}. The statistics that we consider are the colour-magnitude
relation (CMR), due to its important role in cluster identification,
and the cluster galaxy luminosity function (CGLF), which is one of the
simplest statistics that can be made for the population of cluster
galaxies.

In $\S$\ref{sec:galaxy_formation_model} we introduce the galaxy
formation model and describe how we select galaxies in clusters. The
set of observed clusters, which we compare with the model predictions,
are introduced in $\S$\ref{sec:observational_datasets}. In
$\S$\ref{sec:cluster_lf} we compare the model predictions for the
CGLF, to observational estimates between redshifts $z=1.2$ and $z=1.6$
and examine possible factors that might be causing the discrepancy
between the observations and the model predictions. Next, in
$\S$\ref{sec:cluster_cmd}, we compare the model predictions for the
CMR, of cluster galaxies with the observed one. In
$\S$\ref{sec:discussion} we examine the effect of varying selected
model parameters. Finally, we draw our conclusions in
$\S$\ref{sec:conclusions}.

All synthetic and observed magnitudes have been converted to the AB
system.

%%%%%%%%%%%%%%%%%%%%%%%%%%%%%%%%%%%%%%%%%%%%%%%%%%%%%%%%%%%%%%%%%%%%%%%%%%%%%%%%%%%%%%%%%%%%%%%%%%%
% GALAXY FORMATION MODEL
%%%%%%%%%%%%%%%%%%%%%%%%%%%%%%%%%%%%%%%%%%%%%%%%%%%%%%%%%%%%%%%%%%%%%%%%%%%%%%%%%%%%%%%%%%%%%%%%%%%

\section{Galaxy formation model}
\label{sec:galaxy_formation_model}

In this section we describe the galaxy formation model that we employ,
starting with the N-body simulation used ($\S$\ref{sec:MillGas}) and
followed by the semi-analytical model ($\S$\ref{sec:GALFORM}). We then
discuss the dust attenuation treatment used in the model
($\S$\ref{sec:GALFORM_dust}) and explain how we define a galaxy
cluster ($\S$\ref{sec:halo_selection}).

\subsection{N-body simulation}
\label{sec:MillGas}

The cosmological simulation that we use is a revision of the
\emph{Millennium Simulation} \citep{Springel05a}, constructed using
using a cosmology consistent with the 7 year results of the
\emph{Wilkinson Microwave Anisotropy Probe}
\citep[WMAP7,][]{Komatsu11}. The cosmological parameters are: a baryon
matter density $\Omega_{{\rm b}} = 0.0455$, a total matter density
$\Omega_{{\rm m}} = \Omega_{{\rm b}} + \Omega_{{\rm CDM}} = 0.272$, a
dark energy density $\Omega_{{\rm \Lambda}} = 0.728$, a Hubble
constant $H_0 = 100h \,{\rm km\, s}^{-1}\Mpc^{-1}$ where $h = 0.704$,
a primordial scalar spectral index $n_{\rm s}=0.967$ and a fluctuation
amplitude $\sigma_{8}=0.810$. We shall refer to this simulation as the
\emph{MS-W7 Simulation} \citep{Guo13}.

The hierarchical growth of cold dark matter structure is followed from
redshift $z=127$ to the present day, in a cubic volume of size
$500h^{-1}\Mpc$ on a side. Halo merger trees are constructed using
particle and halo data stored at 62 fixed epoch snapshots, which are
spaced approximately logarithmically in expansion factor.  Details
regarding construction of the halo merger trees can be found in
\citet{Merson13} and \citet{Jiang14}. The MS-W7 simulation uses
$2160^3$ particles to represent the matter distribution, with the
requirement that a halo consists of at least 20 particles for it to be
resolved. This corresponds to a halo mass resolution of $M_{{\rm
    halo,lim}} = 1.87\times10^{10}h^{-1}\Msol$, significantly smaller
than expected for the Milky Way's dark matter halo.

\subsection{The GALFORM semi-analytical model}
\label{sec:GALFORM}
To model the star formation and merger history of galaxies we use the
\GALFORM{} semi-analytical model of galaxy formation
\citep{Cole00}. The model populates dark matter halos with galaxies by
using a set of coupled differential equations to determine how the
various baryonic components of galaxies evolve
\citep{Baugh06,Benson10b}.

For the work presented here we use a development of \GALFORM{} that
accounts for the following physical processes: (i) the collapse and
merging of dark matter (DM) halos, (ii) the shock-heating and
radiative cooling of gas inside DM halos, leading to the formation of
galactic discs (iii) quiescent star formation in galactic discs,
explicitly following the atomic and molecular gas components
\citep{Lagos11a,Lagos12}, (iv) feedback as a result of supernovae,
active galactic nuclei \citep{Bower06} and photo-ionisation of the
inter-galactic medium, (v) chemical enrichment of stars and gas, (vi)
dynamical friction driven mergers of galaxies within DM halos, capable
of forming spheroids and triggering starburst events, and (vii) disk
instabilities, which can also trigger starburst events.

Most of the published versions of \GALFORM{} adopt a single
\citet{Kennicutt83} initial mass function (IMF, see \citealt{Baugh05}
for an illustration of using a top-heavy IMF in starbursts) and
updated versions of stellar population synthesis (SPS) models from
\citet{Bruzual93}. (See \citealt{Gonzalez-Perez14} for a comparison of
coupling \GALFORM{} with different SPS models). By combining the star
formation histories of the galaxies with the SPS models, \GALFORM{} is
able to calculate spectral energy distributions (SEDs) for the
galaxies. Absolute magnitudes in a given photometric band can be
obtained by integrating the SED with the corresponding
frequency-dependent filter response curve. (To calculate magnitudes in
the observer-frame, a frequency shift is first applied to the filter
response curve). All magnitudes and colours are total
magnitudes. Apparent magnitudes are calculated using the redshift of
the simulation snapshot to determine the distance modulus
\citep[see][]{Merson13}. Note that the model magnitudes do not include
any photometric uncertainties (see \citealt{Ascaso15} for a discussion
of the impact of photometric errors on the colour-magnitude relation).

\GALFORM{} is able to track the global metallicity for the stars, as
well as the hot and cold gas in the galaxy. Chemical enrichment is
modelled using the instantaneous recycling approximation, with an
effective yield and a recycled fraction that depend upon the choice of
IMF. The yield is modified accordingly by metal ejection and feedback
and hence is a function of the depth of the potential well of the
galaxy. The rate at which gas is ejected from the galaxy due to
supernovae explosions, $\dot{M}_{{\rm eject}}$, is given by,
\begin{equation}
  \dot{M}_{{\rm eject}} = \left (\frac{v_{{\rm hot}}}{v_{{\rm
          disc}}}\right )^{\alpha_{{\rm hot}}}\dot{M}_{\star},
  \label{eqn:sn_feedback}
\end{equation}
where $v_{{\rm disc}}$ is the circular velocity of the galaxy disc at
the half mass radius, $\dot{M}_{\star}$ is the star formation rate and
$\alpha_{{\rm hot}}$ and $v_{{\rm hot}}$ are free parameters that
govern the strength of supernovae feedback.

In \GALFORM{}, feedback due to AGN is implemented in halos that are
undergoing quasi-static cooling, where, at fixed radius, the cooling
time of the hot halo gas, $\tau_{{\rm cool}}$, exceeds the dynamical
free-fall time of the gas, $\tau_{{\rm ff}}$.  Therefore feedback due
to AGN can only occur when the condition,
\begin{equation}
\left.\frac{\tau_{{\rm ff}}}{\tau_{{\rm cool}}}\right |_{r=r_{{\rm
      cool}}}<\alpha_{{\rm cool}},
\label{eq:alpha_cool}
\end{equation}
is satisfied, where $\alpha_{{\rm cool}}$ is a free parameter. This
condition is evaluated at the cooling radius, $r_{{\rm cool}}$, which
is defined, for a halo of a given age, as the radius at which the hot
gas has only just had sufficient time to cool and collapse onto the
galactic disc. Reducing the value of $\alpha_{{\rm cool}}$ raises the
minimum halo mass at which a quasi-static halo is established, thus
allowing star formation to continue for longer in more massive halos.

The free parameters in the \citeauthor{Gonzalez-Perez14} model were
calibrated to reproduce the $\band{b_J}$ and \band{K}-band luminosity
functions at $z=0$ as well as to predict a reasonable evolution for
the rest-frame \band{K}-band and UV luminosity functions. We stress at
this point that this model has not been explicitly constrained using
any observations of high redshift clusters.

Overall, the \GALFORM{} model is able to make predictions for numerous
galaxy properties, including luminosities over a substantial
wavelength range extending from the far-UV through to the
sub-millimetre. However, matching precisely the observed colour
distribution of galaxies in the local Universe remains difficult for
semi-analytical models \citep[e.g.][]{Font08,Gonzalez09,Guo11}.

\subsection{Modelling dust attenuation}
\label{sec:GALFORM_dust}

The attenuation by dust of the starlight from galaxies is modelled in
\GALFORM{} using a physically motivated method based upon the results
of the radiative transfer code of \citet{Ferrara99}. The method
assumes the dust to be distributed in dense molecular clouds embedded
in a diffuse component. In this model the dust attenuation varies
self-consistently with other galaxy properties, such as size, gas mass
and metallicity, which are predicted by \GALFORM{} \citep[see
also][]{Fontanot09a, Fontanot11}.  The \band{V}-band optical depth
when looking face-on through the centre of a galaxy, $\tau_{{\rm
    V}}^0$, is assumed to be,
\begin{equation}
  \tau_{{\rm V}}^0 \propto \frac{M_{{\rm cold}}Z_{{\rm
        cold}}}{r^2_{{\rm disc}}},
  \label{eqn:optical_depth}
\end{equation}
where $M_{{\rm cold}}$ is the mass of cold gas in the galaxy (both
atomic and molecular), $Z_{{\rm cold}}$ is the metallicity of the cold
gas content of the galaxy and $r_{{\rm disc}}$ is the radius of the
galactic disc. Given an extinction law, the \citet{Ferrara99} model
provides dust attenuation factors as a function of wavelength, galaxy
inclination, the ratio of bulge to disc radial dust scale length,
$r_e/h_R$, the ratio of dust to stellar vertical scale heights,
$h_{z{\rm ,\,dust}}/h_{z{\rm ,\,stars}}$, and $\tau_{{\rm V}}^0$ (see
\citealt{Cole00} and \citealt{Gonzalez-Perez14} for further details).

\subsection{Cluster galaxy selection}
\label{sec:halo_selection}
When comparing the semi-analytic predictions to the observational
estimates, we consider clusters of galaxies to be hosted by dark
matter halos with mass, $M_{{\rm halo}}$,
\begin{equation}
M_{{\rm halo}}\geqslant 1.2\times10^{14}h^{-1}\Msol.
\label{eqn:mass_limit}
\end{equation}
This value is chosen as a compromise to ensure that we have a
sufficiently large sample of halos that are massive enough to
adequately represent our set of observed clusters. After imposing the
halo mass limit we are left with 43 halos at $z\sim1.4$, 98 halos at
$z\sim1.2$ and 10 halos at $z\sim1.6$. Cluster member galaxies are
then taken to be those galaxies sharing a common host dark matter
halo. Additionally, we place an aperture of $120^{\arcseconds}$ by
first assuming an observer placed at infinity, viewing the halos along
the Cartesian Z-axis of the simulation box. We use the positions of
galaxies labelled by \GALFORM{} as being central galaxies, which are
located at the centre of mass of the most massive sub-halo of a halo,
as the positions of the halo centres. We then apply the aperture using
the projected distances between the galaxies and the halo centre. The
effect of our choices for the halo mass limit and aperture size, as
well as our method for modelling dust attenuation, are discussed
further in $\S$\ref{sec:galform_predictions}. 

We stress that no colour selection is applied when selecting the
\GALFORM{} cluster galaxies and that, since we know the halo
membership of the galaxies, the model predictions do not include
contamination from foreground or background interlopers. For this work
we are therefore examining the properties of model galaxies that truly
are cluster galaxies, i.e. are hosted by cluster-mass halos selected
according to our threshold of $M_{{\rm halo}}\geqslant
1.2\times10^{14}h^{-1}\Msol$. An assessment of the effect of member
incompleteness and interlopers, where cluster galaxies are selected
according to their colours, is left for future work.

%%%%%%%%%%%%%%%%%%%%%%%%%%%%%%%%%%%%%%%%%%%%%%%%%%%%%%%%%%%%%%%%%%%%%%%%%%%%%%%%%%%%%%%%%%%%%%%%%%%
% OBSERVATIONAL DATA
%%%%%%%%%%%%%%%%%%%%%%%%%%%%%%%%%%%%%%%%%%%%%%%%%%%%%%%%%%%%%%%%%%%%%%%%%%%%%%%%%%%%%%%%%%%%%%%%%%%

\section{Observational datasets}
\label{sec:observational_datasets}

\begin{table*}
\centering
\caption{High redshift observational galaxy cluster datasets used for
  comparison with model predictions. Columns show: (i) Reference and cluster ID,
  (ii) estimated cluster redshift, (iii) photometry available, (iv)
  aperture used to identify cluster members, (v) estimated mass of
  cluster from X-ray measurements, (vi) estimated mass of cluster from
  weak lensing measurements. Cluster mass estimates are provided where
  known, with values converted to units of $10^{14}h^{-1}\Msol$, using
  the values of $h$ specified by the original authors. Note that
  cluster RX J0848+4453 is listed twice.}
\begin{tabular}{| c | c | c | c | c | c |}
\hline
Cluster & Cluster & Photometry & Aperture & X-ray Mass & Lensing Mass\\
& Redshift & & (Arcseconds) &($10^{14}h^{-1}\Msol$)&($10^{14}h^{-1}\Msol$)\\
\hline\hline
\noindent{\underline{\citet{Mei06b}}}&&&&& \\
RX J0849+4452&1.26&${\rm i_{775}}$, ${\rm z_{850}}$&120&$2.0\pm1.0^a$&$3.08^{+0.78}_{-0.63}\,^{b}$\\
RX J0848+4453&1.27&${\rm i_{775}}$, ${\rm z_{850}}$&120&$0.96\pm0.69^a$&$2.2^{+0.7}_{-0.6}\,^b$\\
\hline
{\underline{\citet{Strazzullo06}}}&&&&& \\
RDCS J0910+5422&1.106&${\rm H_{160}}$, ${\rm K}$&73&$2.1\pm1.3^a$&$3.5^{+0.8}_{-0.7}\,^b$\\
RDCS J1252.9-2927&1.237&${\rm H_{160}}$, ${\rm K}$&59&$1.11\pm0.24^a$&$4.8^{+0.9}_{-0.7}\,^b$\\
RX J0848+4453&1.273&${\rm H_{160}}$, ${\rm K}$&47&$0.96\pm0.69^a$&$2.2^{+0.7}_{-0.6}\,^b$\\
\hline
{\underline{\citet{Hilton09}}}&&&&& \\
XMMXCS J2215.9-1738&1.46&$\band{z_{850}}$,\band{J},\band{K},&153$^c$&$1.4^{+0.36}_{-0.42}\,^b$&$3.0^{+2.1}_{-1.2}\,^b$\\
\hline
{\underline{\citet{Strazzullo10}}}&&&&& \\
XMMU J2235-2557&1.39&$\band{z_{850}}$, $\band{H_{160}}$, \band{J}, \band{K},&83&$4.3^{+1.0}_{-0.8}\,^b$&$5.1^{+1.2}_{-1.0}\,^b$\\
\hline
\noindent{\underline{{\citet{Snyder12}}}}&&&&& \\
ISCS J1426.1+3403 & 1.136 & ${\rm H_{160}}$, ${\rm I_{814}}$ &122&-&-\\
ISCS J1426.5+3339 & 1.163 & ${\rm H_{160}}$, ${\rm I_{814}}$ &122&-&-\\
ISCS J1434.5+3427 & 1.243 & ${\rm i_{775}}$, ${\rm H_{160}}$ &120&-&$1.8^{+1.6}_{-0.8}\,^b$ \\
ISCS J1429.3+3437 & 1.262 & ${\rm z_{850}}$, ${\rm H_{160}}$ &120&-&$3.8^{+1.7}_{-1.1}\,^b$ \\
ISCS J1432.6+3436 & 1.349 & ${\rm z_{850}}$, ${\rm H_{160}}$ &119&-&$3.7^{+1.8}_{-1.2}\,^b$ \\
ISCS J1433.8+3325 & 1.369 & ${\rm z_{850}}$, ${\rm H_{160}}$ &119&-&-\\
ISCS J1434.7+3519 & 1.372 & ${\rm z_{850}}$, ${\rm H_{160}}$ &119&-&$2.0^{+2.1}_{-1.0}\,^b$\\
ISCS J1438.1+3414 & 1.413 & ${\rm z_{850}}$, ${\rm H_{160}}$ &119&$2.2^{+3.7}_{-1.0}\,^b$&$2.2^{+1.8}_{-1.0}\,^b$\\
\hline
{\underline{\citet{Fassbender14}}}&&&&& \\
XDCP J0044.0-033&1.58& ${\rm J}$, ${\rm K}$, ${\rm i}$, ${\rm V}$&30&$2.1$&-\\
\hline
\multicolumn{6}{l}{$^a$$M_{500}$ estimate from \citet{Ettori04}} \\
\multicolumn{6}{l}{$^b$$M_{200}$ estimate from \citet{Jee11}} \\
\multicolumn{6}{l}{$^c$Based upon maximum radial distance quoted in Table 1
  of \citet{Hilton09}} \\
\end{tabular}
\label{tab:obs_datasets}
\end{table*}

Here we briefly introduce the observational datasets against which we
will compare the model predictions.

Observations of clusters at high redshift, $z>1$, are still limited to
of the order of a few tens of clusters, often with masses exceeding
$10^{14}h^{-1}\Msol$. Here, we consider a compilation of clusters at
redshifts between $z\sim1.2$ and $z\sim1.6$, for which multi-band
photometry is available. In particular, we focus on clusters for which
photometry is available such that they could be identified by the
\Euclid{} mission. The clusters that we consider are listed in
Table~\ref{tab:obs_datasets}.

\begin{figure}
  \centering
  \includegraphics[width=0.46\textwidth]{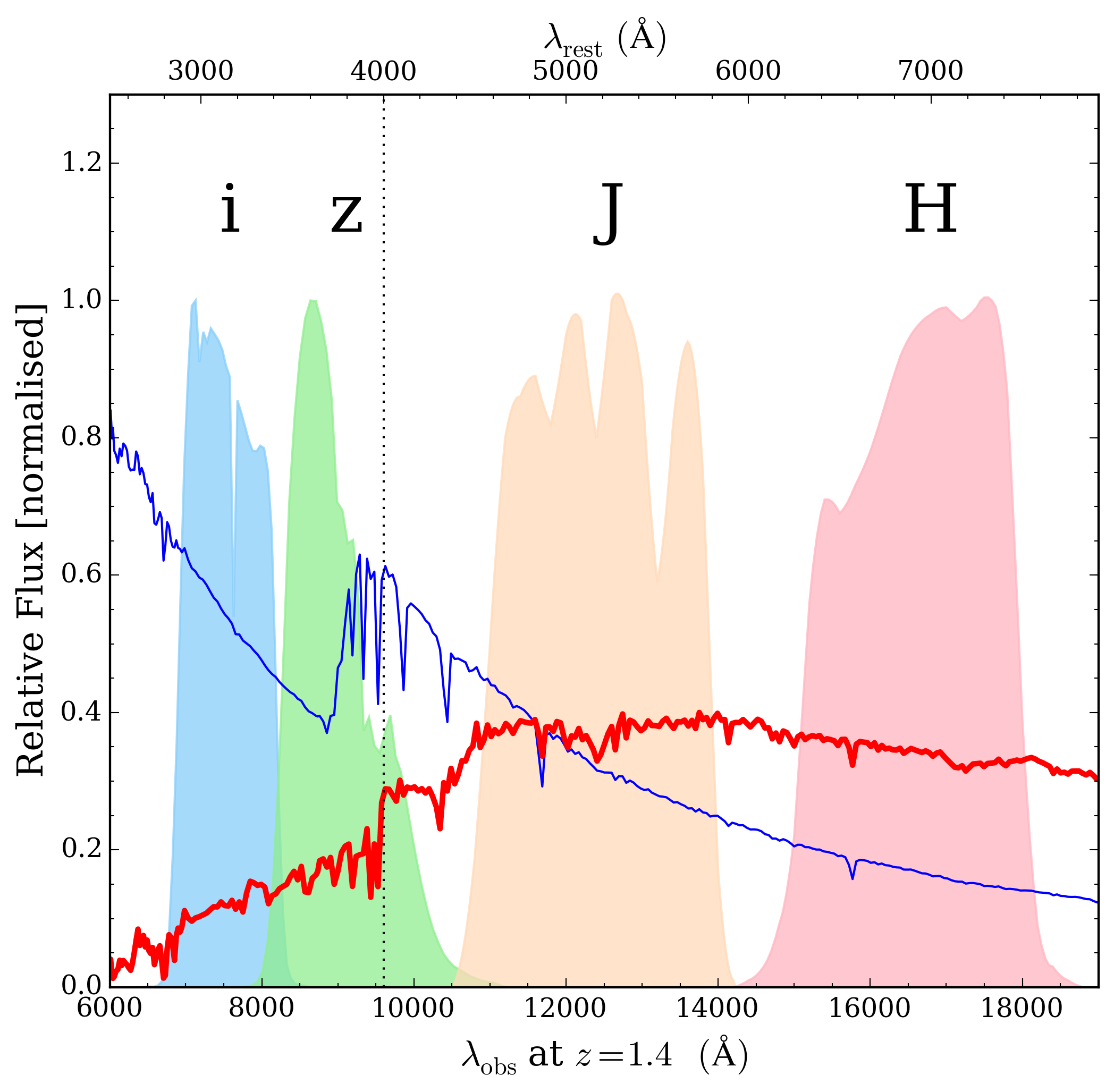}
  \caption{Examples of intrinsic spectra for two synthetic galaxies at
    redshift $z=1.4$, over-plotted with transmission profiles for the
    ${\rm i}$, ${\rm z}$, ${\rm J}$ and ${\rm H}$ bands, as shown by
    the shaded regions. No dust attenuation is considered. Both of the
    galaxies were assumed to undergo a single instantaneous burst of
    star formation, one $300{\rm Myr}$ ago (thin blue line) and the
    other $3{\rm Gyr}$ ago (thick red line), prior to $z=1.4$. (Both
    galaxies are assumed to have a metallicity of $Z=0.008$). The
    spectra were generated using the {\tt PEGASE.2} code
    \protect\citep{Fioc99}, assuming a \protect\cite{Kennicutt83}
    IMF. The dotted, vertical line indicates the rest-frame wavelength
    of $4000{\rm \AA}$.}
  \label{fig:example_spectra}  
\end{figure}

Typically, the colours used to examine cluster galaxies are chosen
such that the pairs of photometric bands bracket the redshifted
$4000{\rm \AA}$ break, the strength of which is typically used as a
proxy for the age of a galaxy and as a way of distinguishing passively
evolving galaxies from those that are undergoing star formation when
combined with other indexes \citep{Kauffmann03a,Kriek06a,Kriek11}, as
illustrated in Fig.~\ref{fig:example_spectra} for two synthetic
galaxies at $z=1.4$. At $z\sim1.2$ the $4000{\rm \AA}$ break is
shifted to $8800{\rm \AA}$, which lies between the \band{i}-band and
\band{z}-band, whereas at $z\sim1.4$ the $4000{\rm \AA}$ break, now
shifted to $9600{\rm \AA}$, is bracketed by the \band{z}-band and
\band{J}-band. At $z\sim1.6$ the $4000{\rm \AA}$ break is shifted to
$10400{\rm \AA}$ and is still bracketed by the \band{z}-band and
\band{J}-band.

For the majority of the clusters, mass estimates are available. In
Table~\ref{tab:obs_datasets} we provide $M_{200}$ and $M_{500}$ mass
estimates from \citet{Ettori04} and \citet{Jee11}. In many cases the
mass estimates from gravitational lensing are larger than the mass
estimates based upon the X-ray emission from the cluster, which is
likely to be due to projection effects in the lensing
estimate. However, X-ray mass estimates are also uncertain due to the
limitations in our understanding of the conditions in X-ray gas
\citep[e.g.][]{Angulo12}.

In the remainder of this section we provide further details on each of
the galaxy cluster datasets.

\begin{itemize}
\item \textit{\citet{Mei06b}}: \citeauthor{Mei06b} present
  $\band{i_{775}}$ (F775W) and $\band{z_{850}}$ (F850LP) observations
  of the clusters RX J0849+4452 and RX J0848+4453, which together make
  up the Lynx Supercluster, located at $z\sim1.2$. Applying the colour
  selection $0.8<\left (\band{i_{775}}-\band{z_{850}}\right )<1.1$,
  the flux selection $21<\band{z_{850}}<24$ and an aperture selection
  of $120^{\arcseconds}$, left 40 galaxies, of which 14 are confirmed
  cluster members and 26 are cluster member candidates. Galaxy colours
  were measured within the effective radii of the
  galaxies. \citeauthor{Mei06b} assumed reddening due to dust to be
  described by a \citet{Calzetti00} extinction law of ${\rm
    E(B-V)}=0.027$ with $A_{\band{i_{775}}}=0.054$ and
  $A_{\band{z_{850}}}=0.040$.
\item \textit{\citet{Strazzullo06}}: \citeauthor{Strazzullo06} present
  $\band{K_s}$-band imaging for the clusters RDCS J0910+5422
  ($\band{K_s}<21.5$) and RDCS J1252.9-2927 ($\band{K_s}<24.5$), as
  well as $\band{H_{160}}$ (F160W) imaging for RX J0848+4453
  ($\band{H_{160}}<25$). They present an estimate for the CGLF for
  each cluster, as well as a composite estimate for all three of the
  clusters. The apertures applied are listed in
  Table~\ref{tab:obs_datasets}.
\item \textit{\citet{Hilton09}}: \citeauthor{Hilton09} present
  \band{J}-and $\band{K_s}$-band photometry, along with
  $\band{z_{850}}-\band{J}$ and $\band{z_{850}}-\band{K_s}$ colour
  information, for 64 galaxies selected as members of the cluster
  XMMXCS J2215.9-1738. The radial distance of the galaxies relative to
  the cluster X-ray source position extends out to $922\,{\rm
    kpc}$. Spectroscopic redshifts are obtained for 24 of these
  galaxies. The photometry was corrected for Galactic extinction using
  the dust emission maps of \citet{Schlegel98}.
\item \textit{\citet{Strazzullo10}}: \citeauthor{Strazzullo10} present
  multi-wavelength data for the cluster XMMU J2235-2557, one of the
  most massive virialised structures found beyond $z\sim1$. They
  present estimates for the CGLF of the cluster in $\band{z_{850}}$,
  $\band{H_{160}}$ and $\band{K_s}$, with $10\sigma$ completeness
  limits of $\band{z_{775}}<25.3$, $\band{H_{160}}<25$ and
  $\band{K_s}<23$. They also apply an aperture and select only those
  galaxies within $83^{\arcseconds}$ of the cluster centre. The
  photometry was corrected for Galactic extinction using the dust
  emission maps of \citet{Schlegel98}. \citeauthor{Strazzullo10} argue
  that for their dataset the population of brightest galaxies in the
  inner region of the cluster can be considered to be quite well
  established, with about seventy per cent of these galaxies having
  measured spectroscopic redshifts.
\item \textit{\citet{Snyder12}}: \citeauthor{Snyder12} present Hubble
  Space Telescope follow up observations of clusters selected from the
  \textit{Spitzer}/IRAC Shallow Cluster Survey
  \citep[ISCS,][]{Eisenhardt08}. They provide photometry for these
  clusters in the $\band{H_{160}}$-band and one of the
  $\band{I_{814}}$, $\band{i_{775}}$ or $\band{z_{850}}$ bands. They
  report that the detection catalogues for each cluster are more than
  90 per cent complete for $\band{H_{160}}<23.5$. To identify cluster
  galaxies lying on the red sequence, they subtract from the colour of
  each galaxy a fiducial evolved CMR model for the Coma cluster and
  apply a selection based upon the residual. Reddening due to dust is
  assumed to be small; approximately ${\rm E(B-V)}\lesssim0.06$.
\item \textit{\citet{Fassbender14}}: \citeauthor{Fassbender14} present
  VLT/HAWK-I \band{J} and $\band{K_s}$-band observations of the
  cluster XDCP J0044.0-033, complemented with the \band{V} and
  \band{i}-bands from Subaru archival imaging. They report that their
  observations are 100 per cent complete down to
  $\band{J}\lesssim23.9$ and $\band{K_s}\lesssim23.8$. To maximise the
  signal-to-noise ratio of the cluster members to the interlopers,
  \citeauthor{Fassbender14} apply an aperture of
  $30^{\arcseconds}$. They estimate that the galaxies within this
  radius are 90 per cent cluster-associated.
\end{itemize}

%%%%%%%%%%%%%%%%%%%%%%%%%%%%%%%%%%%%%%%%%%%%%%%%%%%%%%%%%%%%%%%%%%%%%%%%%%%%%%%%%%%%%%%%%%%%%%%%%%%
% CLUSTER GALAXY LUMINOSITY FUNCTION
%%%%%%%%%%%%%%%%%%%%%%%%%%%%%%%%%%%%%%%%%%%%%%%%%%%%%%%%%%%%%%%%%%%%%%%%%%%%%%%%%%%%%%%%%%%%%%%%%%%

\section{The cluster galaxy luminosity function}
\label{sec:cluster_lf}

\begin{figure*}
  \centering
  \includegraphics[width=0.94\textwidth]{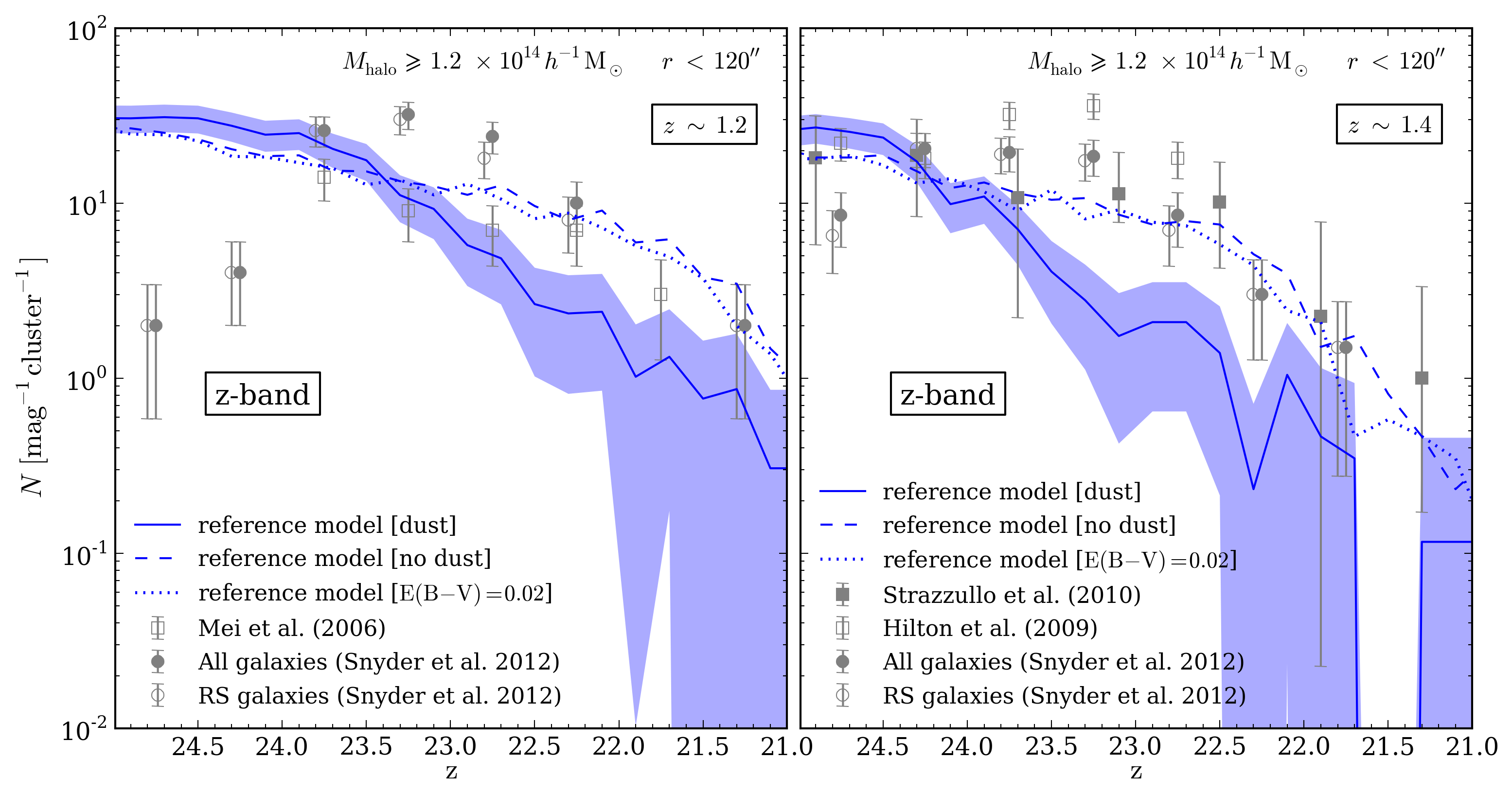}
  \caption{\band{z}-band cluster galaxy luminosity functions (CGLF) at
    redshifts $z\sim1.2$ (left) and $z\sim1.4$ (right). The blue,
    solid line shows the predicted CGLF for the reference model, with
    dust attenuation. The shaded region indicates the Poisson
    uncertainty on this prediction. The dashed line shows the
    prediction for the reference model with no dust attenuation
    applied, i.e. using the fluxes intrinsic to the galaxies. The
    dotted line shows the prediction for the reference model when a
    dust attenuation similar to a \protect\citet{Calzetti00}
    extinction law with $\mathrm{E}(\mathrm{B}-\mathrm{V})=0.02$ is
    assumed. All model predictions correspond to the CGLF for all
    cluster galaxies within an aperture of $120^{\arcseconds}$, hosted
    by dark matter halos of $M_{{\rm halo}}\geqslant
    1.2\times10^{14}h^{-1}\Msol$.  Observational estimates of the
    luminosity functions are indicated by various data points.}
  \label{fig:cglf_zband}
\end{figure*}

\begin{figure*}
  \centering
  \includegraphics[width=0.94\textwidth]{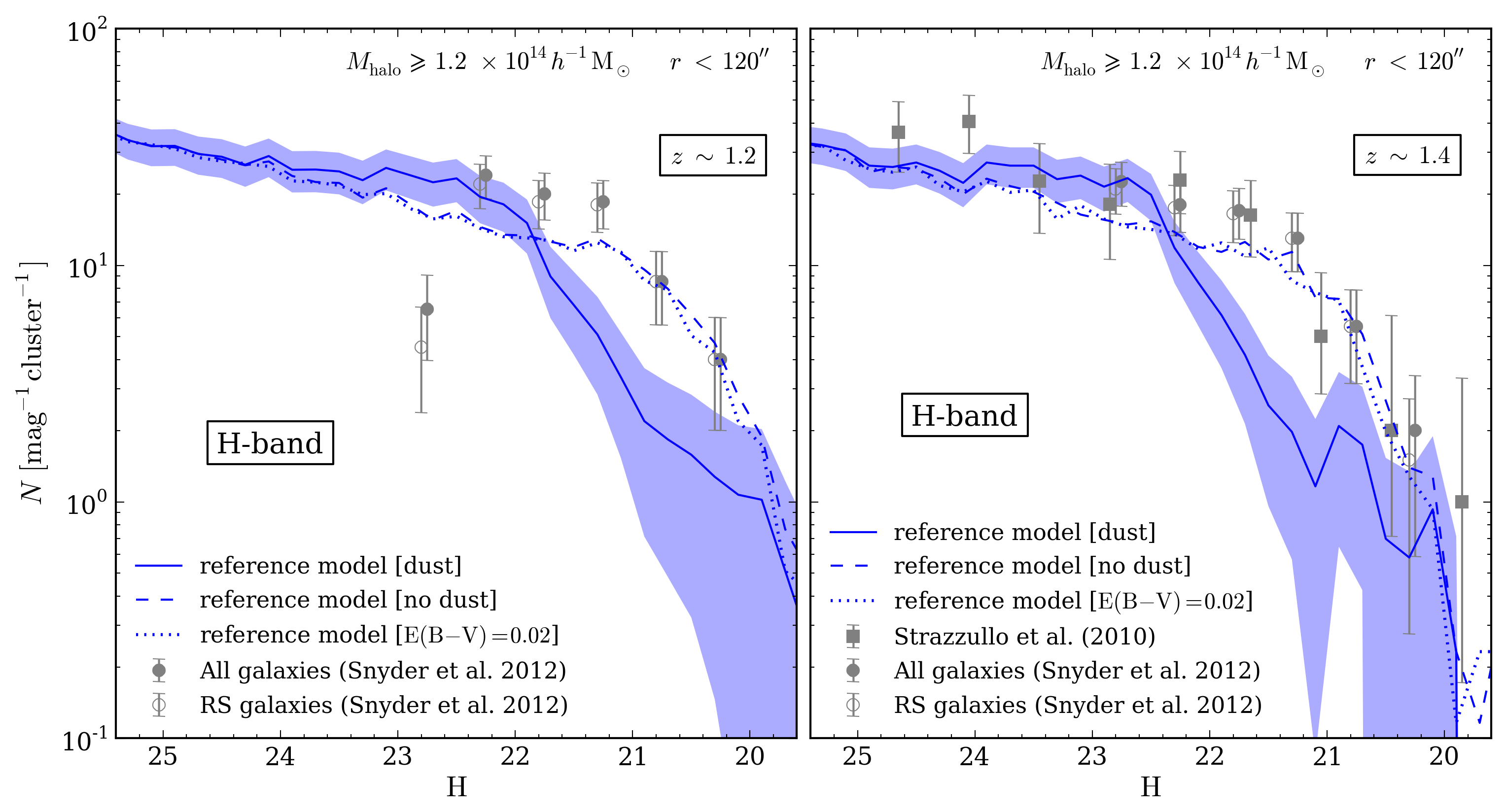}
  \caption{\band{H}-band cluster galaxy luminosity functions (CGLF) at
    redshifts $z\sim1.2$ (left) and $z\sim1.4$ (right). All model
    predictions correspond to the CGLF for all cluster galaxies, with
    galaxies selected in the same way as in
    Fig.\ref{fig:cglf_zband}. Observational estimates of the
    luminosity functions are indicated by various data points.}
  \label{fig:cglf_hband}
\end{figure*}

\begin{figure*}
  \centering
  \includegraphics[width=0.97\textwidth]{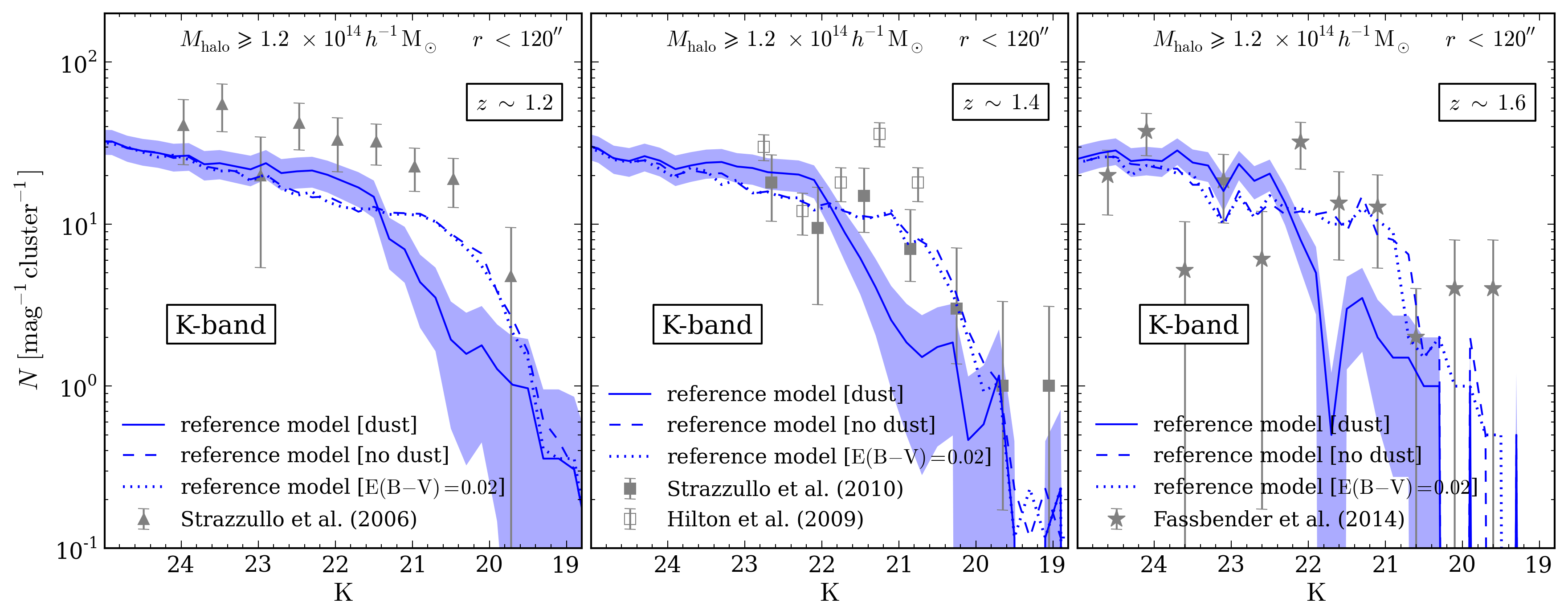}
  \caption{\band{K}-band cluster galaxy luminosity functions (CGLF) at
    redshifts $z\sim1.2$ (left) and $z\sim1.4$ (middle) and $z\sim1.6$
    (right). All model predictions correspond to the CGLF for all
    cluster galaxies, with galaxies selected in the same way as in
    Fig.\ref{fig:cglf_zband}. Observational estimates of the
    luminosity functions are indicated by various data points.}
  \label{fig:cglf_kband}
\end{figure*}

In this section we compare observational estimates for the CGLF, which
describes the number of galaxies per cluster as a function of apparent
magnitude, with the predictions from our reference model. We stress
again that for the model predictions, knowledge of the halo membership
of the galaxies means that we can simply select cluster galaxies using
a halo mass selection and that no further colour selections are
imposed. The observational estimates and semi-analytical predictions
for the \band{z}-band CGLF are shown in Fig.~\ref{fig:cglf_zband}, the
\band{H}-band CGLF in Fig.~\ref{fig:cglf_hband} and the \band{K}-band
CGLF in Fig.~\ref{fig:cglf_kband}. The observational estimates are
shown by the various data points, whilst the model predictions are
indicated by the solid lines (with shaded regions indicating the
Poisson uncertainties).

\subsection{Observational estimates}
\label{sec:obs_estimates}
In Fig.~\ref{fig:cglf_zband}, Fig.~\ref{fig:cglf_hband} and
Fig.~\ref{fig:cglf_kband} we show observational estimates for the CGLF
from \citet{Strazzullo06}, \citet{Strazzullo10} \citet{Fassbender14},
as well as our estimates based upon the datasets of \citet{Mei06b},
\citet{Hilton09} and \citet{Snyder12}.

\citet{Strazzullo06} and \citet{Strazzullo10} estimate the CGLF for
their galaxy cluster sample by using a reference field, down to an
equivalent photometric depth or deeper, to statistically remove the
contribution from background galaxies, which is known to bias CGLF
estimates \citep{Andreon05}, particularly at bright magnitudes where
the statistics are typically quite poor. The CGLF is estimated by
subtracting the counts in the reference field (normalised to the solid
angle of the cluster) from the counts in the cluster field. The
uncertainties on the CGLFs estimated by \citet{Strazzullo06} and
\citet{Strazzullo10} are Poisson, with both clusters and possible
background field interlopers summed in quadrature. The limits on the
excess counts are determined based upon the upper and lower limits on
the number of sources with a spectroscopic redshift or a photometric
redshift within $3\sigma$ of the cluster
redshift. \citet{Strazzullo06} argue that the background contamination
from lensed galaxies magnified by the cluster itself is small. They
conclude that their estimate for the CGLF is consistent with previous
determinations at similar or lower
redshifts. \citeauthor{Fassbender14} adopt the same selection
procedure as \citet{Strazzullo06} when estimating the $z\sim1.6$
\band{K}-band CGLF.

For the datasets of \citeauthor{Mei06b}, \citeauthor{Hilton09} and
\citeauthor{Snyder12} we have made simple estimates for the CGLF by
counting all of the galaxies regarded as being cluster members by the
original authors. From the \citeauthor{Mei06b} data we estimate the
\band{z}-band CGLF at $z\sim1.2$. From the \citeauthor{Hilton09} data
we estimate the \band{z} and \band{K}-band CGLFs at $z\sim1.4$. From
the \citeauthor{Snyder12} data we estimate the \band{z} and
\band{H}-band CGLFs at $z\sim1.2$ and $z\sim1.4$. We apply no further
selection beyond those placed originally by the authors and provide
simple Poisson uncertainties on these counts. As such, our estimated
uncertainties on these observational luminosity functions may well be
under-estimated and our CGLF estimates may be biased by membership
incompleteness.

At $z\sim1.4$, in the $\band{z}$ and $\band{H}$ bands, we compare our
estimates for the CGLF to the estimates from \citet{Strazzullo10},
though these are for a slightly more massive cluster and adopt a
smaller aperture when selecting the member galaxies (as we demonstrate
in $\S$\ref{sec:aperture_size}, changing the aperture size has a
greater impact at the faintest magnitudes). We find that our CGLF
estimates from the \citeauthor{Snyder12} data are consistent within
error with the estimates from \citeauthor{Strazzullo10}. When
constructing their dataset, \citeauthor{Snyder12} identified galaxies
as cluster members based upon the likelihood of the galaxy lying on
the cluster RS. For a subset of the galaxies these authors were able
to confirm that they indeed are on the RS of the clusters.  In
Fig.~\ref{fig:cglf_zband} and Fig.~\ref{fig:cglf_hband} we plot the
\citeauthor{Snyder12} estimate for the CGLF using only the confirmed
RS galaxies and can see that this is in excellent agreement with the
estimate using the full \citeauthor{Snyder12} dataset, suggesting that
our CGLF estimates are not being significantly biased by possible
interloping non-member galaxies.

At $z\sim1.2$, there are no existing estimates of the CGLF in the
\band{H} or \band{z}-bands against which to compare ours. In the these
bands our estimates from the \citeauthor{Snyder12} dataset show a
bright-end fall off similar in shape to the CGLF estimates at
$z\sim1.4$. At the faintest magnitudes, however, the CGLFs show a
down-turn. \citeauthor{Snyder12} report $5\sigma$ detection limits of
$26.0-26.3$ in the \band{z}-band and $24.4-24.8$ in the \band{H}-band,
with $90\%$ completeness at $\band{H}\sim23.5$. It is possible
therefore that the observed down-turns may be a result of
incompleteness. When determining their samples of cluster galaxies,
\citeauthor{Snyder12} place a selection limit of $\band{H}\sim22-23$,
determined for each cluster by evolving the characteristic brightness
of the Coma cluster to the redshift of each cluster. This selection
may also be contributing to the down-turns.

In the \band{z}-band, our estimate for the CGLF from the
\citeauthor{Mei06b} shows good agreement with the
\citeauthor{Snyder12} estimate for magnitudes bright-wards of
$\band{z}\sim22.2$. Faint-wards of this value, however, the estimates
diverge. We note that these estimates are based upon galaxies from
only one or two clusters, so discrepancies are indeed possible due to
cosmic variance across the cluster populations. Another possible cause
could be due to the $\band{i}-\band{z}$ colour selection that
\citeauthor{Mei06b} place in order to identify early-type galaxies. We
note, however, that the uncertainties on our CGLF estimates are simple
Poisson errors and so may well be underestimates of the true
uncertainties. If this is the case, then it is possible that our
estimates for these two datasets are consistent within error.

Our estimates for the \band{z} and \band{K}-band CGLFs from the
\citeauthor{Hilton09} dataset are generally consistent with the other
estimates presented, though there are one or two bins for which the
counts are higher than the other estimates, for example at
$\band{z}\sim23.2$ and $\band{K}\sim21.2$. Just under two-thirds of
galaxies in the \citeauthor{Hilton09} data have photometric redshifts
only and so this variation could be caused by interlopers. In
addition, variation due to cosmic variance between CGLF measurements
for different clusters would be expected.

\subsection{Semi-analytical predictions}
\label{sec:galform_predictions}
The predictions for the CGLF from our reference \GALFORM{} model are
shown in Fig.~\ref{fig:cglf_zband}, Fig.~\ref{fig:cglf_hband} and
Fig.~\ref{fig:cglf_kband} by the solid blue lines, with shaded regions
indicating the Poisson uncertainties. These predictions correspond to
the CGLF for \emph{all} cluster galaxies, i.e. all galaxies in halos
with mass $M_{{\rm halo}}\geqslant 1.2\times10^{14}h^{-1}\Msol$,
within a projected aperture of $r<120^{\arcseconds}$.

There is a large discrepancy between the model predictions and the
observational estimates around the knee of the CGLF. This discrepancy
is most obvious in the \band{H} and \band{K}-band CGLF comparisons,
though it is still apparent in the \band{z}-band CGLF, particularly at
$z\sim1.4$. In the brightest and faintest magnitude bins, however, the
model predictions for the CGLF are broadly consistent with the
observational estimates.

In the \band{H} and \band{K}-bands, the model predictions for the CGLF
show a faint-end slope that is quite flat, in agreement with the
faint-end slope seen in the observations. In the \band{K}-band,
however, the normalisation of the slope in the model predictions is
lower than that of the observations, particularly at $z\sim1.2$.

\begin{figure}
  \centering
  \includegraphics[width=0.46\textwidth]{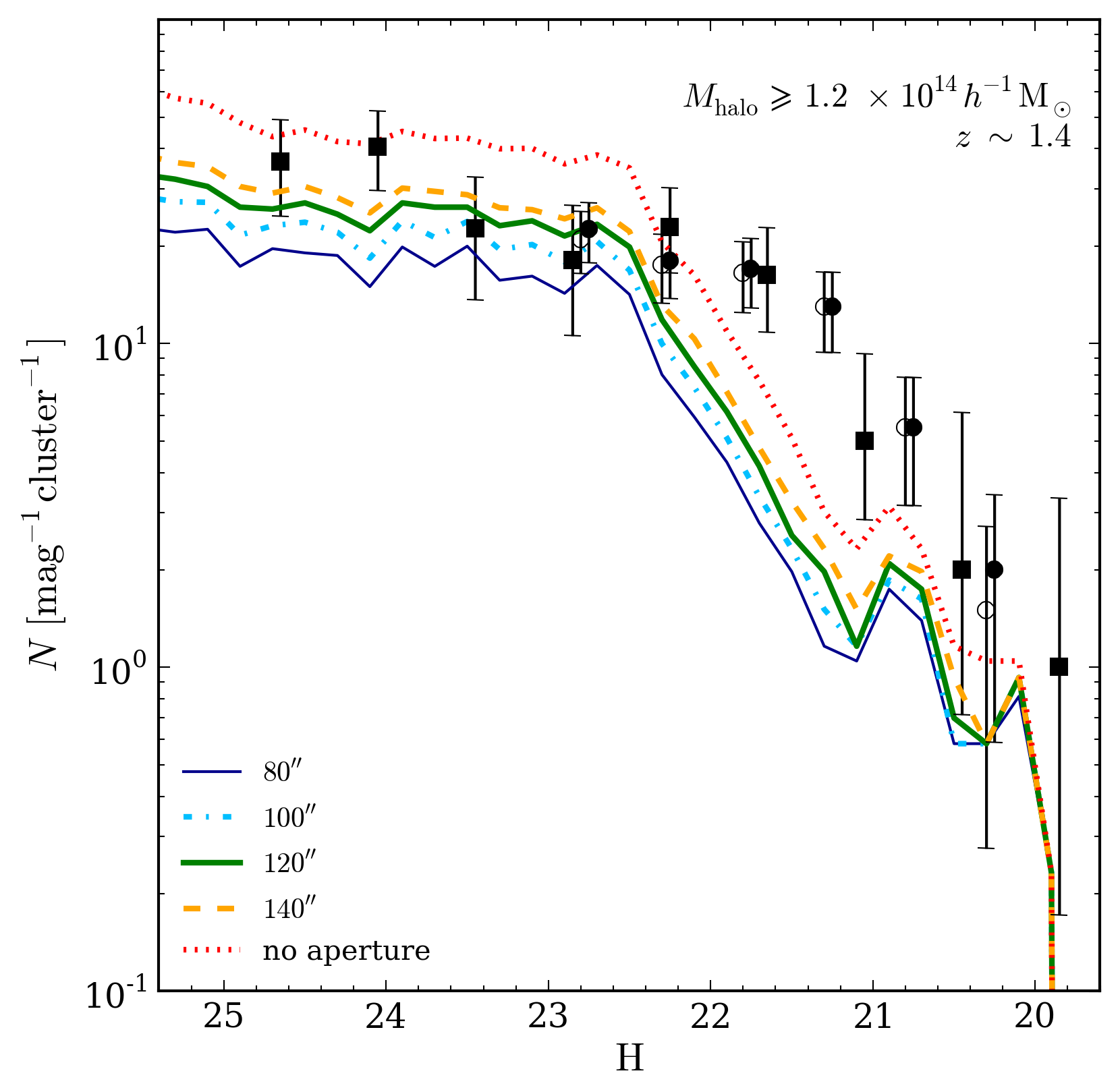}
  \caption{\band{H}-band CGLFs at $z\sim1.4$ as predicted by
    \GALFORM{} assuming different apertures for selecting the member
    galaxies of each halo, as indicated in the legend. For each choice
    of aperture a halo mass limit of $M_{{\rm halo}}\geqslant
    1.2\times10^{14}h^{-1}\Msol$ was adopted. The data points are the
    same as in the right-hand panel of Fig.~\ref{fig:cglf_hband}.}
  \label{fig:varying_aperture}
\end{figure}

We now consider three factors that could affect our
comparison with cluster data, particularly our estimation of the
CGLF. These are: the size of the aperture we apply to the halo
($\S$\ref{sec:aperture_size}), our choice of halo mass limit applied
to the model ($\S$\ref{sec:halo_mass_limit}) and our modelling of
attenuation due to dust ($\S$\ref{sec:dust_correction}).

\subsubsection{Aperture size}
\label{sec:aperture_size}
We select the \GALFORM{} cluster galaxies using an aperture size of
$r<120^{\arcseconds}$, which is consistent with the apertures placed
by \citet{Mei06b} and \citet{Snyder12}. Given the cosmology used in
the MS-W7 simulation, an aperture of $120^{\arcseconds}$ corresponds
approximately to a distance of $500h^{-1}{\rm kpc}$ at
$z\sim1.4$. Varying the aperture size will change the number of
satellite galaxies that are included. We must therefore examine how a
change in aperture size affects our estimates of the CGLF.

In Fig.~\ref{fig:varying_aperture} we plot the predictions for the
reference model for the \band{H}-band CGLF at $z\sim1.4$ when keeping
the halo mass limit fixed and allowing the aperture size to vary. The
variation of the aperture size induces a change in the abundance of
cluster members, particularly for the faintest magnitude bins where a
change in the aperture size will lead to different numbers of faint
satellite galaxies being selected. This change in the abundance,
typically within a factor of two, is consistent with observational
uncertainties. All of the apertures we consider correspond to
distances smaller than the virial radius of a typical cluster-sized
dark matter halo and so remove the most distant satellites from the
comparison such that we are comparing only the cores of the
clusters. At brighter magnitudes, $\band{H}\lesssim21.5$, the change
in the abundance becomes smaller due to the increasing number of
bright, central galaxies, which will always be selected. If we remove
the aperture altogether then we see an increase of approximately a
factor of two in the abundance of cluster members with
$\band{H}\gtrsim21.5$, as all of the most distant satellites are now
included (shown by the red line in Fig.~\ref{fig:varying_aperture}).

From Fig.~\ref{fig:varying_aperture} we can see that an aperture of
$120^{\arcseconds}$ is consistent with many of the aperture choices in
Table~\ref{tab:obs_datasets} and provides a suitable match to the
counts just faint-wards of the break in the CGLF.

\subsubsection{Halo mass limits}
\label{sec:halo_mass_limit}

We select cluster galaxies in \GALFORM{} as galaxies hosted by halos
above a threshold mass of $1.2\times10^{14}h^{-1}\Msol$. Using a fixed
aperture of $120^{\arcseconds}$ we find that variation of the halo
mass limit between $1.0\times10^{14}h^{-1}\Msol$ and
$1.8\times10^{14}h^{-1}\Msol$ produces a negligible change in
predicted \band{H}-band CGLF at $z\sim1.4$. (Examination of larger
halo masses is not possible due to the limited volume of the MS-W7
simulation.) We find that the choice of halo mass limit has a
negligible effect on the CGLF.

\subsubsection{Attenuation due to dust}
\label{sec:dust_correction}
The dust content of a galaxy can have a drastic effect upon the
observed colour of the galaxy. In Figs.~\ref{fig:cglf_zband},
\ref{fig:cglf_hband} and \ref{fig:cglf_kband} we compare the model
CGLF, both with and without dust attenuation, with the
observations. In the reference model dust attenuation has a large
impact around the knee of the CGLF. At the knee of the CGLF the
reference model is about a factor of 4.5 below the observations. The
observed CGLF can be reproduced if instead we consider a model without
dust attenuation. However, such a model predicts an unrealistic
luminosity function at $z=0$, as we shall discuss later.

\begin{figure}
  \centering
  \includegraphics[width=0.46\textwidth]{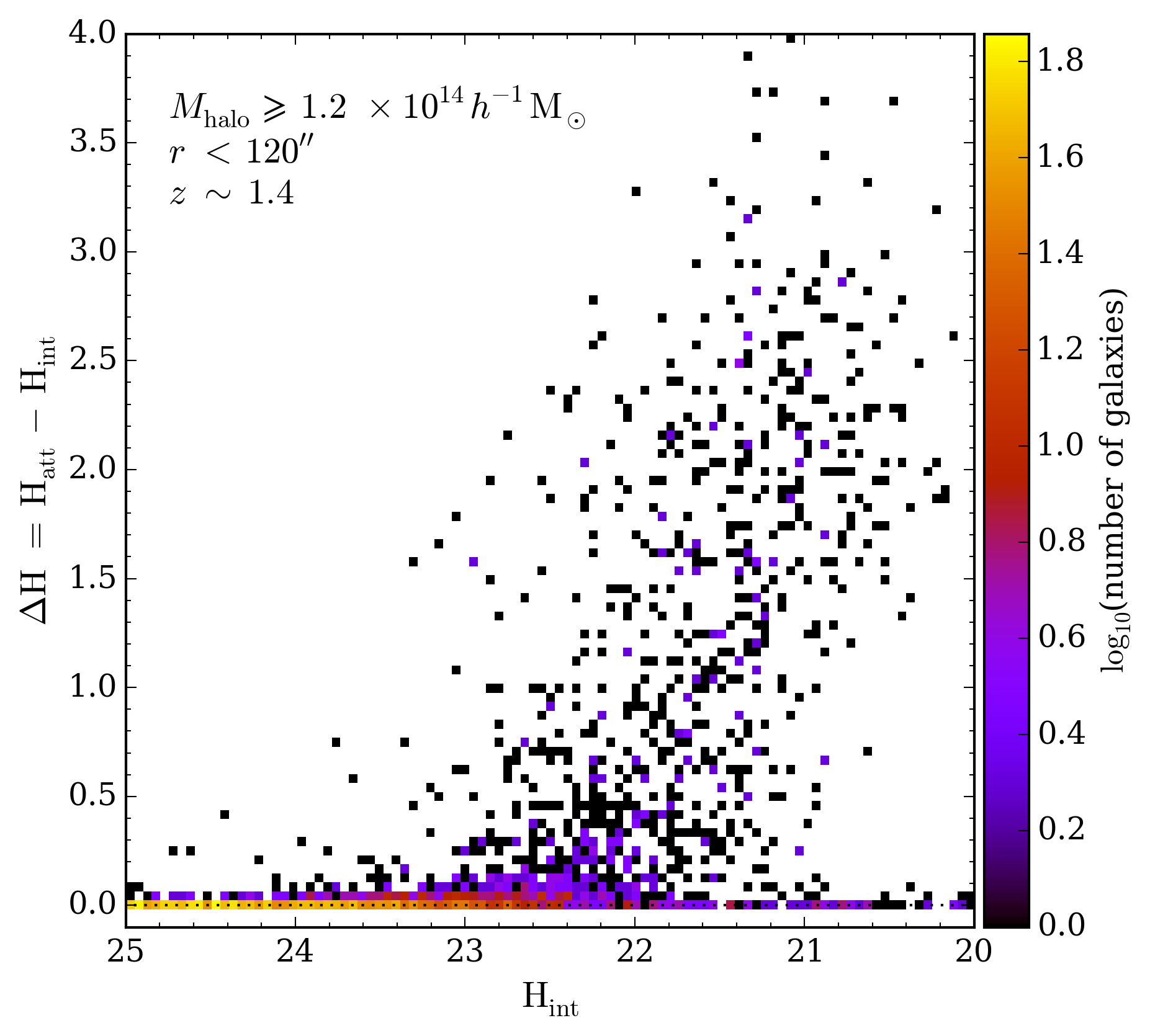}
  \caption{Difference in \GALFORM{} \band{H}-band attenuated
    magnitudes, $\band{H_{att}}$, and \band{H}-band intrinsic
    magnitudes, $\band{H_{int}}$, as a function of intrinsic
    magnitude. (The default dust attenuation calculation is
    assumed). The colour bar indicates the number of galaxies in each
    pixel. The redshift, as well as the halo mass and aperture used to
    select the \GALFORM{} galaxies, are shown in top right of the
    panel.}
  \label{fig:dust_attenuation}
\end{figure}

To help understand this result, we compare the intrinsic and
attenuated magnitudes of the \GALFORM{} galaxies in our reference
model. We define the difference,
\begin{equation}
\Delta \band{M} = \band{M_{att.}} - \band{M_{int.}},
\label{eq:dust_diff}
\end{equation}
where $\band{M_{int}}$ is the intrinsic, dust-free magnitude of a
\GALFORM{} galaxy and $\band{M_{att.}}$ is the magnitude of this
galaxy attenuated using the dust model described in
$\S$\ref{sec:GALFORM_dust}. In Fig.~\ref{fig:dust_attenuation} we plot
the difference between these two magnitudes as a function of
\band{H}-band intrinsic magnitude. The majority of galaxies display
very little dust attenuation and so have a negligible difference
between their intrinsic and attenuated magnitudes.

There are, however, a small number of galaxies, with intrinsic
magnitudes between $21<\band{H}_{{\rm int.}}<23$, that display dust
attenuation larger than one magnitude. If we examine again the
\band{H}-band CGLF at $z\sim1.4$ (right-hand panel, of
Fig.~\ref{fig:cglf_hband}), we can see that this magnitude range
corresponds approximately to the knee of the CGLF, where the maximum
discrepancy between the model and the observations occurs. For
magnitudes outside this range, the attenuation is minimal. Hence, the
size of the attenuation applied in our reference model is not constant
with magnitude, unlike, for example, the simple \citet{Calzetti00}
fitting formula that is often applied to low redshift star-forming
galaxies. With our reference attenuation calculation, the large dust
attenuation in the brightest galaxies causes these galaxies to be
shifted out of the brightest magnitude bins and to pile up in the
magnitude bins faint-wards of the characteristic magnitude. Although
these galaxies constitute only of the order $5-10$ per cent of
galaxies with $\band{H_{att.}}<25$ for example, the rapidly declining
number of galaxies in the brightest bins means that this magnitude
shift has a significant impact upon the CGLF.

As a final demonstration that these highly dust attenuated galaxies
are the cause of the discrepancy between the model CGLF and the
observations, we calculate again the CGLF for our reference model but
now assume a weaker dust attenuation, similar to a \citet{Calzetti00}
extinction law with $\mathrm{E}(\mathrm{B}-\mathrm{V})=0.02$. The
resulting CGLF predictions are shown as blue dotted lines in
Figs.~\ref{fig:cglf_zband}, \ref{fig:cglf_hband} and
\ref{fig:cglf_kband}. In each instance, we see that weakening the
attenuation in this way leads to a much better agreement between the
model and the observations.

%%%%%%%%%%%%%%%%%%%%%%%%%%%%%%%%%%%%%%%%%%%%%%%%%%%%%%%%%%%%%%%%%%%%%%%%%%%%%%%%%%%%%%%%%%%%%%%%%%%
% CLUSTER COLOUR MAGNITUDE RELATION
%%%%%%%%%%%%%%%%%%%%%%%%%%%%%%%%%%%%%%%%%%%%%%%%%%%%%%%%%%%%%%%%%%%%%%%%%%%%%%%%%%%%%%%%%%%%%%%%%%%

\section{The cluster colour-magnitude relation}
\label{sec:cluster_cmd}

Having examined the abundance of semi-analytical cluster galaxies, we
now consider their colours. Specifically, we consider the
colour-magnitude relation (CMR). In the CMR the red sequence (RS) is
often used observationally to detect clusters as it is expected to be
dominated by early-type cluster members. It is desirable to examine
the model predictions for the RS as any discrepancy between the model
and observations could, for example, impact upon the calibration of
cluster-finding algorithms for next generation surveys such as Euclid.

\begin{figure*}
  \centering
  \includegraphics[width=0.96\textwidth]{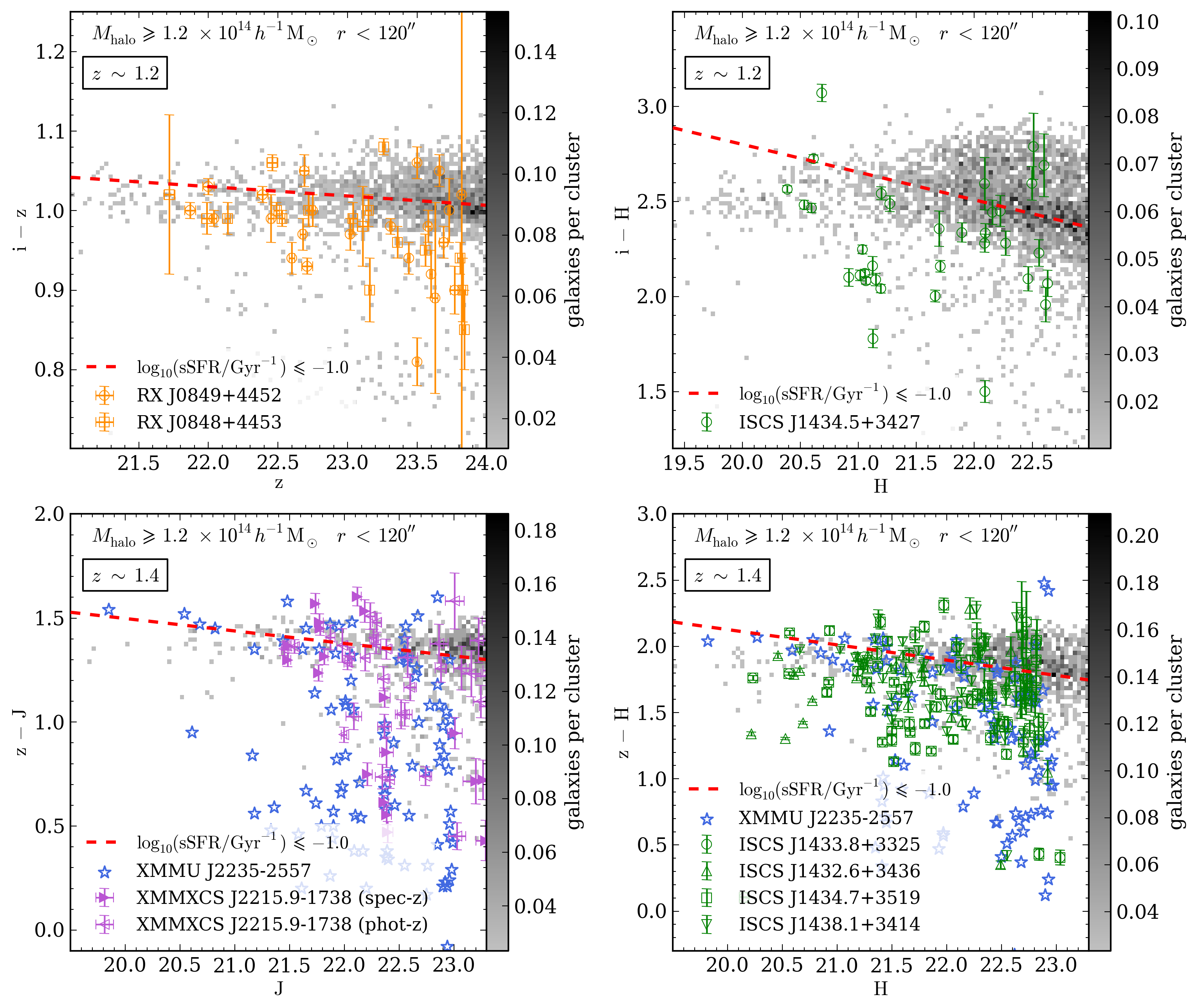}
  \caption{\GALFORM{} colour-magnitude relations at redshifts
    $z\sim1.2$ (upper panels) and $z\sim1.4$ (lower panels). The
    observed cluster galaxies are shown by the various data
    points. Greyscale pixels show the number of cluster galaxies per
    cluster as predicted by the fiducial \GALFORM{} model. The red
    dashed lines show linear fits to the red sequence predicted by
    \GALFORM{}. The halo mass lower limit and aperture, $r$, used to
    select the \GALFORM{} galaxies are shown in the label at the top
    of each panel.}
  \label{fig:galform_cmr}
\end{figure*}

\subsection{Red sequence fitting}
\label{sec:rs_fitting}

\input{rsfits_table_accepted.tex}

To describe the location of the model RS, we use linear
regression\footnote{To determine the optimised value for the slope and
  zero-point we provide the list of galaxy magnitudes and colours to
  the \texttt{curve\_fit} function in the Scientific Python library,
  {\tt scipy} (\url{http://www.scipy.org/}).} to fit the optimised
slope, $s$, and zero-point, $c_{22.5}$, for the relation,
\begin{equation}
  c = s(\band{m}-22.5)+c_{22.5},
  \label{eqn:RS_linear_fit}
\end{equation}
where $\band{m}$ is the galaxy magnitude, e.g. \band{H}, and $c$ is
the corresponding colour, e.g. $\band{z}-\band{H}$.  Since the RS is
expected to be dominated by passively evolving galaxies and since
\GALFORM{} is able to provide us with values for the star formation
rates of galaxies, as well as which halos they belong to, we are able
to fit a RS using just those galaxies that are passively evolving and
reside in halos above our specified halo mass threshold. Therefore, we
stress that this is not meant to mimic observational methods for
determining the RS, but simply to provide an estimate of the RS
predicted by the model. To determine the passively evolving cluster
galaxies, we apply a cut in instantaneous specific star formation rate
(sSFR) and fit to only those galaxies that satisfy
$\log_{10}(\mathrm{sSFR}/\mathrm{Gyr}^{-1})\leqslant-1$, which
provides a reasonable distinction between actively star-forming and
passively evolving galaxies in observational data
\citep[e.g.][]{Williams09} and hydrodynamical simulations
\citep[e.g.][]{Romeo08,Furlong15}. We therefore regard the fit to the
passively evolving galaxies as our measure of the true RS predicted by
the model, i.e. the RS of those cluster galaxies that are truly `red
and dead'. Note that in addition to the sSFR cut, we also select only
those galaxies brighter than $25^{\mathrm{th}}$ magnitude in the
appropriate band (\band{z}, \band{J}, \band{H} or \band{K}).

The fits to the passively evolved cluster galaxies provide an
excellent description of the model RS down to $25^{\mathrm{th}}$
magnitude, as evident from Figs.~\ref{fig:galform_cmr} and
\ref{fig:cmr_JK} where the fits to the passive RS are shown by the
red, dashed lines. In these figures the greyscale pixels show the
distribution of all cluster galaxies in the model, passive and
star-forming, normalised by the number of clusters (i.e. the number of
halos above the halo mass threshold). In each case the distribution is
dominated by a clear and well-defined RS, especially at faint
magnitudes, with very little indication of a blue cloud.

At bright magnitudes the model RS appears to display a prominent plume
of galaxies with colours extending redwards above the RS. This plume
is visible in many of the colour-spaces that we consider, in
particular $\band{i}-\band{H}$, $\band{z}-\band{H}$ and
$\band{J}-\band{K}$. As we shall see in $\S$\ref{sec:trends}, the
galaxies in the plume are star-forming and so are not identified by
our sSFR selection. As such, they are not included in the fitting and
do not bias the fits to the slope or the zero-point. In contrast,
simply fitting the RS to a straight-forward flux-selected sample,
i.e. without selecting just the passively evolving galaxies, would
lead to biased fits. The results of the fits, i.e. the slope and
intercept, as well as their corresponding uncertainties, are provided
in Table~\ref{tab:RS_fits}.

\subsection{Comparison with observed clusters}
\label{sec:rs_obs_comp}

\begin{figure}
  \centering
  \includegraphics[width=0.46\textwidth]{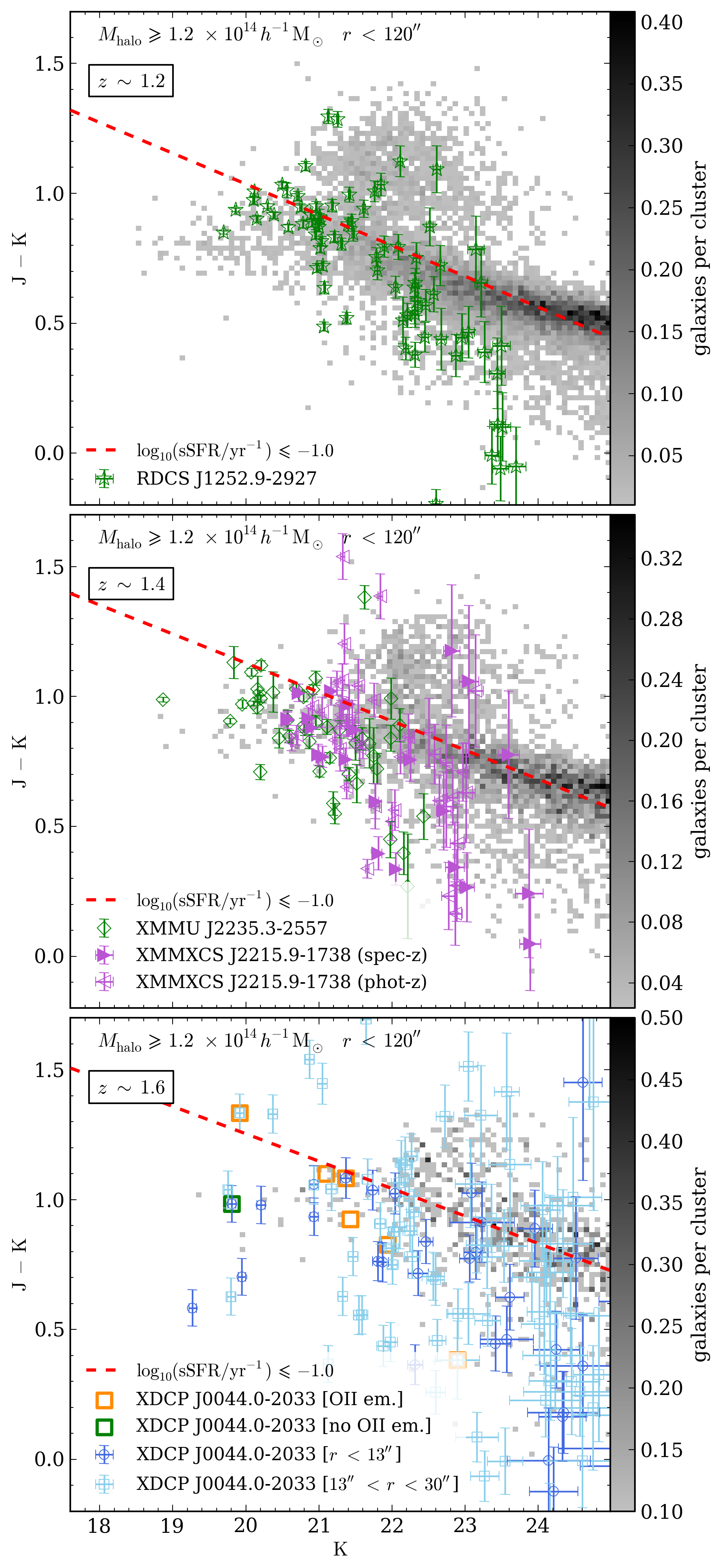}
  \caption{$\band{J}-\band{K}$ vs. $\band{K}$ colour-magnitude
    relations for cluster galaxies at redshifts $z\sim1.2$ (top
    panel), $z\sim1.4$ (middle panel) and $z\sim1.6$ (bottom
    panel). The colours of the observed galaxies are shown by the
    various data points. The observations at $z\sim1.6$, from
    \protect\citet{Fassbender14}, have been split according to
    distance $r$ from the estimated cluster centre: galaxies within
    $r<13^{\arcseconds}$ (which typically have a 75\% membership
    probability) and galaxies with
    $13^{\arcseconds}<r<30^{\arcseconds}$ (which typically have a 50\%
    membership probability). Galaxies that have been spectroscopically
    confirmed are highlighted as being passive (no OII detection) or
    star-forming (OII detection). Note that three of the
    spectroscopically confirmed galaxies lie at
    $r>30^{\arcseconds}$. Cluster data from \protect\citet{Hilton09}
    is split into galaxies with spectroscopic redshifts (spec-z) and
    those without spectroscopic redshifts (photo-z). As before, the
    greyscale pixels show the prediction for the fiducial \GALFORM{}
    model and the red, dashed line shows the linear fit to the
    predicted red sequence. The halo mass and aperture used to select
    the \GALFORM{} galaxies are shown at the top of each panel.}
  \label{fig:cmr_JK}
\end{figure}

We now compare the \GALFORM{} prediction for the cluster CMR with
observational measurements at $z>1$. In Figs.~\ref{fig:galform_cmr}
and \ref{fig:cmr_JK} we show the colours of the observed cluster
galaxies on top of the \GALFORM{} predictions, shown by the greyscale
pixels. In addition, we show as a red, dashed line the fit to the
passively evolving RS. The distribution of \GALFORM{} cluster galaxies
appears to be qualitatively in good agreement with the observations
for each of the colour-magnitude spaces that we consider. In each
instance the model RS is qualitatively in good agreement with
observations. For all cases in Figs.~\ref{fig:galform_cmr} and
\ref{fig:cmr_JK}, the scatter in the distribution of colours from
\GALFORM{} is consistent with or smaller than the spread in the
observations, though we must recall that no photometric errors are
included in the model predictions.

Comparing our estimates for the slope and zero-point (see
Eq.~\ref{eqn:RS_linear_fit}) of the model RS with fits available for
the observed clusters, we find a reasonable agreement for the
longer-wavelength colours, particularly for the zero-point
estimates. For example, our fit to the RS in $\band{z}-\band{J}$ are
consistent with the fit from \citet{Hilton09} who estimated a slope of
$-0.049\pm0.062$ and a zero-point of $1.335\pm0.046$. For the clusters
ISCS J1433.8+3325, ISCS J1432.6+3436, ISCS J1434.7+3519 and ISCS
J1438.1+3414 \citeauthor{Snyder12} determine a range of values for the
$\band{z}-\band{H}$ zero-point. Taking the mean of these values and
adding the uncertainties in quadrature gives an estimate of
$1.78\pm0.09$, which is consistent within $1\sigma$ with our
$\band{z}-\band{H}$ fit. For $\band{i}-\band{z}$, our fit to the RS
zero-point are close to the fit of \citet{Mei06b}, who found a
zero-point of $0.99\pm0.01$. Our fit to the slope, however, are
shallower than their fit of $-0.031\pm0.012$.

\subsection{Trends in galaxy properties}
\label{sec:trends}

\begin{figure}
  \centering
  \includegraphics[width=0.46\textwidth]{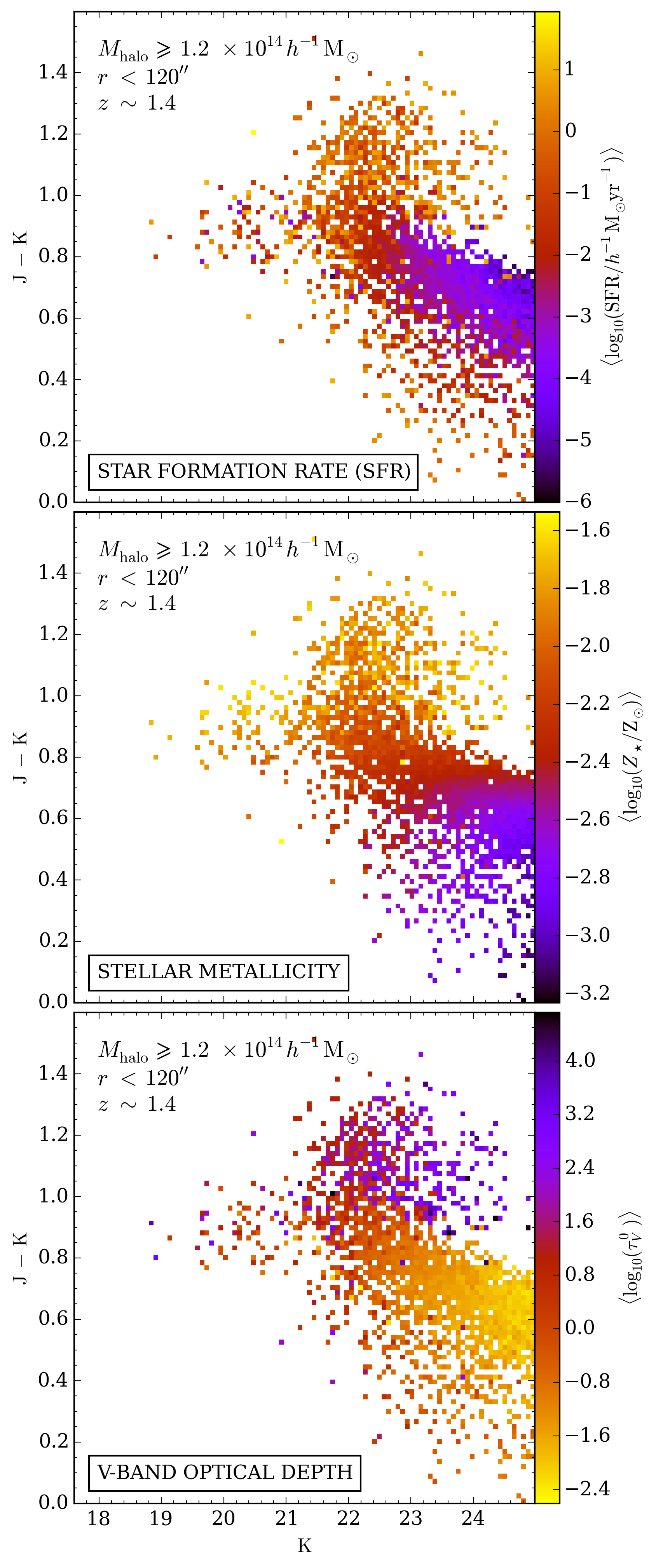}
  \caption{Distribution of various galaxy properties across the
    $\band{J}-\band{K}$ vs. $\band{K}$ colour-magnitude space at
    $z\sim1.4$. The panels, which show the distribution of
    star-formation rate, stellar metallicity and combined half-mass
    radius of the disc and bulge of the galaxies, are labelled
    accordingly. Each pixel is coloured according to the median value
    for the galaxies that lie in that pixel, with the colour scale
    shown at the right of each panel. The halo mass and aperture used
    to select the galaxies are shown at the top of each panel.}
  \label{fig:cmr_JK_properties}
\end{figure}

We conclude our analysis of the cluster RS by examining the properties
of the galaxies in the CMR as predicted by our reference \GALFORM{}
model. In Fig.~\ref{fig:cmr_JK_properties} we show a selection of the
studied properties for $\band{J}-\band{K}$. In these plots the
colour-magnitude space has been divided into pixels and the colour
maps show the median value of a particular property for all of the
cluster galaxies that fall in that particular pixel. We can see
immediately a clear trend in galaxy properties along the RS. The upper
panel of Fig.~\ref{fig:cmr_JK_properties} shows that the brightest and
reddest galaxies typically having higher SFRs than faint red
galaxies. The middle panel also shows that the brightest and reddest
galaxies are also typically the most metal rich. Given also that these
bright, red galaxies also have the largest stellar masses, this result
suggests a positive correlation between stellar mass, metallicity and
SFR. The existence of such a correlation between models and
observations has been debated in the literature
\citep[e.g.][]{Mannucci10,Magrini12,Lilly13,Obreja14}.

In addition, the plume of bright, red galaxies that we have previously
commented on is clearly evident in
Fig.~\ref{fig:cmr_JK_properties}. The galaxies in the plume are
revealed to typically be highly star forming (with star formation
rates in excess of $1-10\,h^{-1}\Msolyr{}$), have reservoirs of cold
gas much larger than other galaxies on the RS, be heavily attenuated
by dust and have metallicities that are richer than the other galaxies
on the RS. The lower panel of Fig.~\ref{fig:cmr_JK_properties} shows
the galaxies in the plume have very large optical depths, much larger
than other cluster galaxies on the RS. The large dust attenuation for
the galaxies in the plume would suggest that these are the same
galaxies that are responsible for the discrepancy between the observed
CGLF and the model prediction, though the exact cause of their large
attenuation is not immediately clear. If we examine the predicted RS
when we apply a weak Calzetti-like attenuation to the intrinsic
magnitudes of the galaxies, then we find that the plume is
removed. This happens at the expense of making the majority of bright
galaxies up to one magnitude bluer, leading to a bluewards break in
the bright RS. This suggests that the plume is an artefact of the
model and that simply reducing the strength of the dust attenuation in
the model is not an acceptable solution. Instead, this perhaps hints
at an additional underlying problem in the model.

Although we suspect that the plume of star-forming galaxies in the
model RS is an artefact, the scatter in the observations hints at the
existence of some observed cluster galaxies with colours as red as
those in the plume. Several observations of our sample of high
redshift clusters also suggest ongoing star formation activity. For
the cluster XDCP J0044.0-2033, \citet{Fassbender14} were able to
obtain spectroscopic redshifts for a few galaxies with significant
[OII] emission, which is often taken as an indicator for ongoing star
formation. One or two of these galaxies, which we have highlighted in
the bottom panel of Fig.~\ref{fig:cmr_JK}, have very red
$\band{J}-\band{K}$ colours suggesting such red galaxies might indeed
be found in clusters in reality, though the large spread in the
observations makes this unclear. \citet{Demarco07} presented
spectroscopy for the cluster of \citet{Strazzullo06} and measured
[OII] emission lines in 38 cluster members. They estimated that the
SFRs of those galaxies are in the range $0.5-2\,\Msolyr{}$, with the
median SFR being $\approx 0.7\,\Msolyr$. Similarly,
\citet{Strazzullo10} reported SFRs of a similar magnitude for cluster
galaxies from photometry, rest-frame FUV. All of these SFR tracers
suggest inferred SFRs that are not as high as the SFRs predicted by
the model for those galaxies in the plume above the RS, although these
SFR tracers can be heavily obscured. However measurements of near-IR
spectroscopy targeting the ${\rm H\alpha}$ line, which is a more
reliable SFR tracer, and IR photometry from Spitzer and Herschel hint
to a higher SFR. For example, \citet{Valentino15} find the ${\rm
  H\alpha}$ luminosity of cluster CL J1449+0856, at $z = 1.99$, to be
significantly higher than measurements in the field at the same
epoch. They attribute this to an enhanced specific star formation rate
in the cluster. In addition, based upon measurements of the
$\band{H}\alpha$ emission from galaxies in the \citet{Snyder12}
clusters, \citet{Zeimann13} infer unobscured SFRs of up to
$200\,\Msolyr{}$ for galaxies right down in the cluster cores. SFR
measurements of the same clusters from Spitzer $24\mu{\rm m}$
observations, as well as Herschel SPIRE data, find similarly high
SFRs, with the SFR in clusters increasing rapidly with redshift
\citep{Brodwin13,Alberts14}.

%%%%%%%%%%%%%%%%%%%%%%%%%%%%%%%%%%%%%%%%%%%%%%%%%%%%%%%%%%%%%%%%%%%%%%%%%%%%%%%%%%%%%%%%%%%%%%%%%%%
% DISCUSSION
%%%%%%%%%%%%%%%%%%%%%%%%%%%%%%%%%%%%%%%%%%%%%%%%%%%%%%%%%%%%%%%%%%%%%%%%%%%%%%%%%%%%%%%%%%%%%%%%%%%

\section{Discussion}
\label{sec:discussion}

We have seen in $\S$\ref{sec:dust_correction} that the dust
attenuation calculation used in our reference \GALFORM{} model
predicts a large dust attenuation for cluster galaxies that leads to
the CGLF predicted by the model being inconsistent with
observations. In addition, in $\S$\ref{sec:cluster_cmd} we have seen
that this large attenuation also produces a plume of very red galaxies
above the RS, which could potentially bias predictions for the RS. We
have seen that whilst applying a weaker dust attenuation removes the
discrepancy between the observed CGLF and the model predictions, this
leads to many of the galaxies in the RS being made too blue. Here we
examine the cause of the large dust attenuation that produces the
tension between the model predictions and observations.

\subsection{Galaxy stellar mass-size relation}
\label{sec:mass_size_relation}

\begin{figure}
  \centering
  \includegraphics[width=0.46\textwidth]{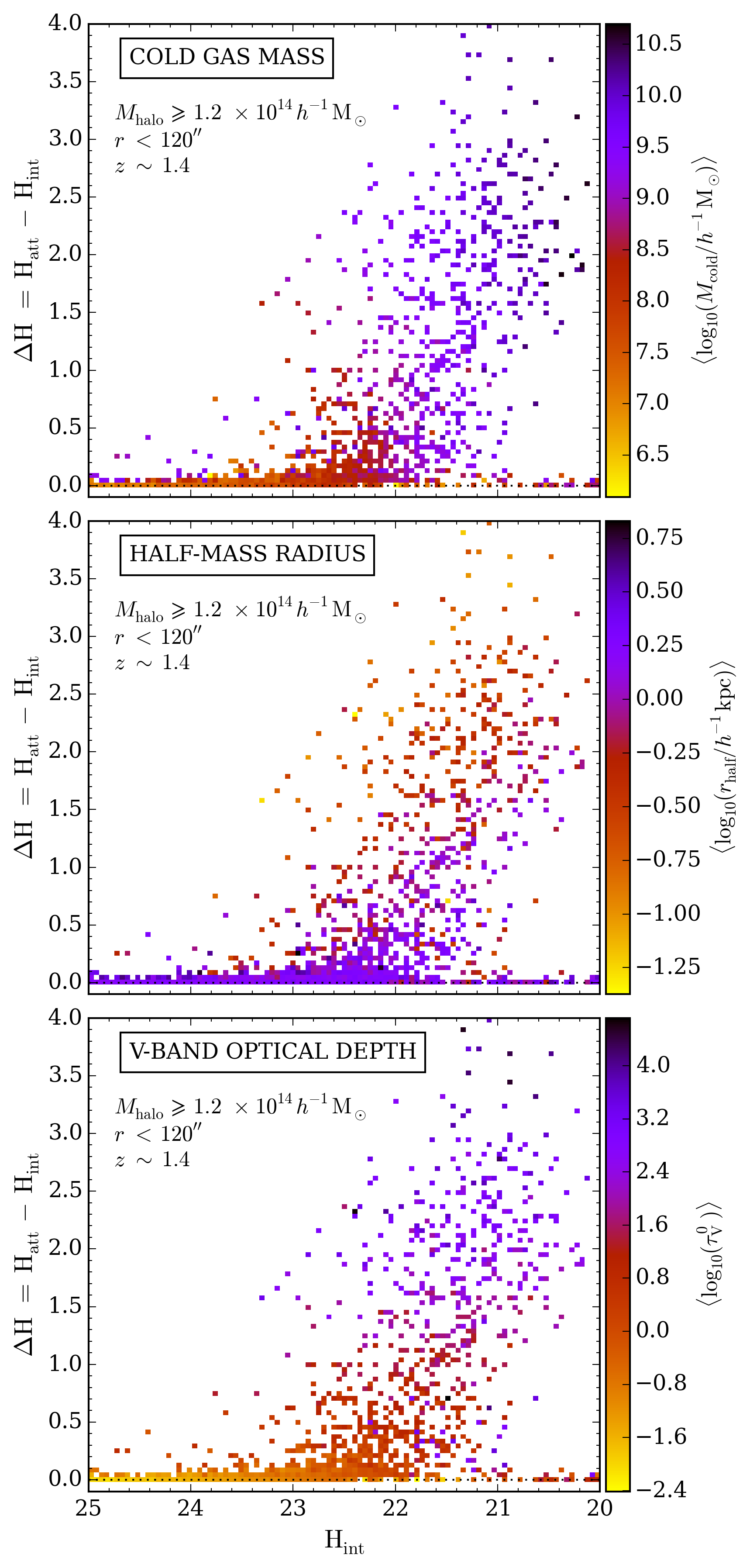}
  \caption{Difference between the \band{H}-band attenuated magnitudes
    and intrinsic magnitudes, as in Fig.~\ref{fig:dust_attenuation},
    but now showing the correlation with selected galaxy properties:
    cold (HI, HII and He) gas mass, half-mass radius and \band{V}-band
    optical depth (through centre of galaxy when face-on). The pixels
    are coloured according to the median value of the galaxy property
    for the galaxies in that pixel as shown by the key on the right
    side of each panel. The redshift, as well as the halo mass and
    aperture used to select the \GALFORM{} galaxies, are shown in each
    panel.}
  \label{fig:dust_attenuation_medians}
\end{figure}

\begin{figure}
  \centering
  \includegraphics[width=0.46\textwidth]{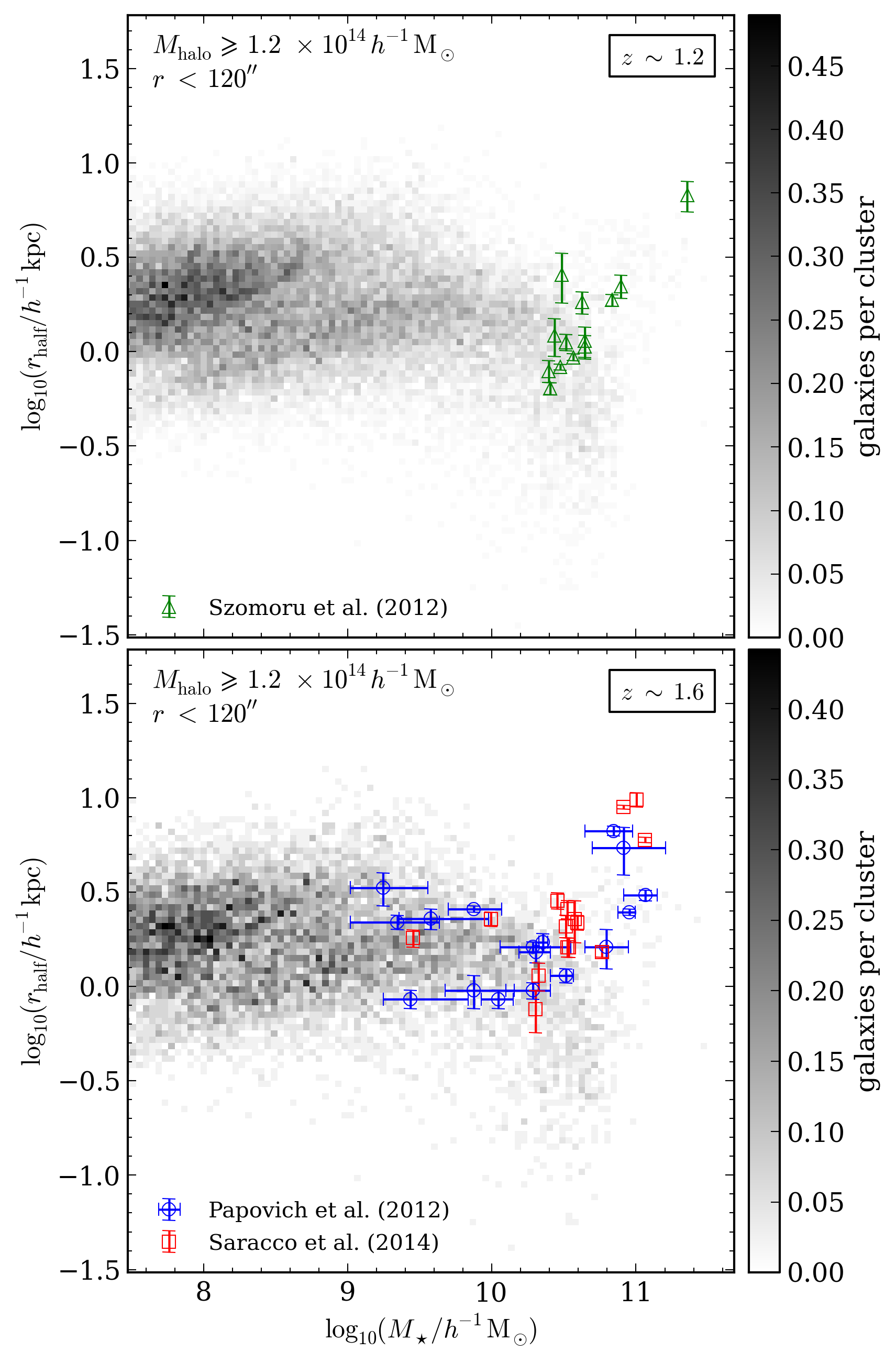}
  \caption{Comparison of the mass-size relation predicted by
    \GALFORM{}, with observational measurements for high-redshift
    early-type galaxies made by \protect\citet{Papovich12,Szomoru12}
    and \protect\citet{Saracco14}. The observational estimates were
    converted from effective radius, $r_e$, to half-mass radius,
    $r_{{\rm half}}$, using the conversion, $r_{{\rm
        half}}=1.35r_e$. The upper panel shows the model predictions
    and observational measurements at $z\sim1.2$, whilst the lower
    panel shows the predictions and measurements at $z\sim1.6$. The
    halo mass and aperture used to select the semi-analytical cluster
    galaxies are shown in the top -left-hand corner of each panel.}
  \label{fig:size_mass}
\end{figure}

From Eq.~\ref{eqn:optical_depth} we can see that the dust attenuation
predicted in our reference model is affected by the predicted galaxy
sizes, cold gas masses and the assumed distribution of dust with
respect to stars. We have tested that changing the distribution of
dust with respect to that of the galaxy stars has little impact on the
CGLF. In Fig.~\ref{fig:dust_attenuation_medians} we show how selected
galaxy properties change with as a function of the difference between
the attenuated and intrinsic \band{H}-band magnitudes of the
\GALFORM{} galaxies. We see from the bottom panel that galaxies with a
large magnitude difference have a larger optical depth We can also see
that galaxies with large attenuation typically have larger reservoirs
of cold gas and smaller radii.  Both the predicted galaxy sizes and
cold gas masses are fundamental predictions from the model that are
directly related to the modelling of the cooling of gas and feedback
processes (see \citealt{Cole00} for details of how galaxy sizes are
calculated in the model). As such, directly modifying these properties
is a complex procedure and beyond the scope of this paper. We do,
however, in $\S$\ref{sec:galform_parameters} briefly explore how
varying selected parameters of the reference model affects the
predictions for the CGLF.

Some observations at high redshift appear to be consistent with high
redshift galaxies having a small dust attenuation
\citep[e.g.][]{Meyers12}. If we assume therefore that there is indeed
negligible dust attenuation in high redshift cluster galaxies, then
from Eq.~\ref{eqn:optical_depth} we can see that too large an optical
depth might indicate that the predicted sizes of such galaxies are too
small, the amount of cold gas in the galaxies is too high or the
metallicity of the cold gas is too high, or a combination of all
three.

\begin{figure}
  \centering
  \includegraphics[width=0.46\textwidth]{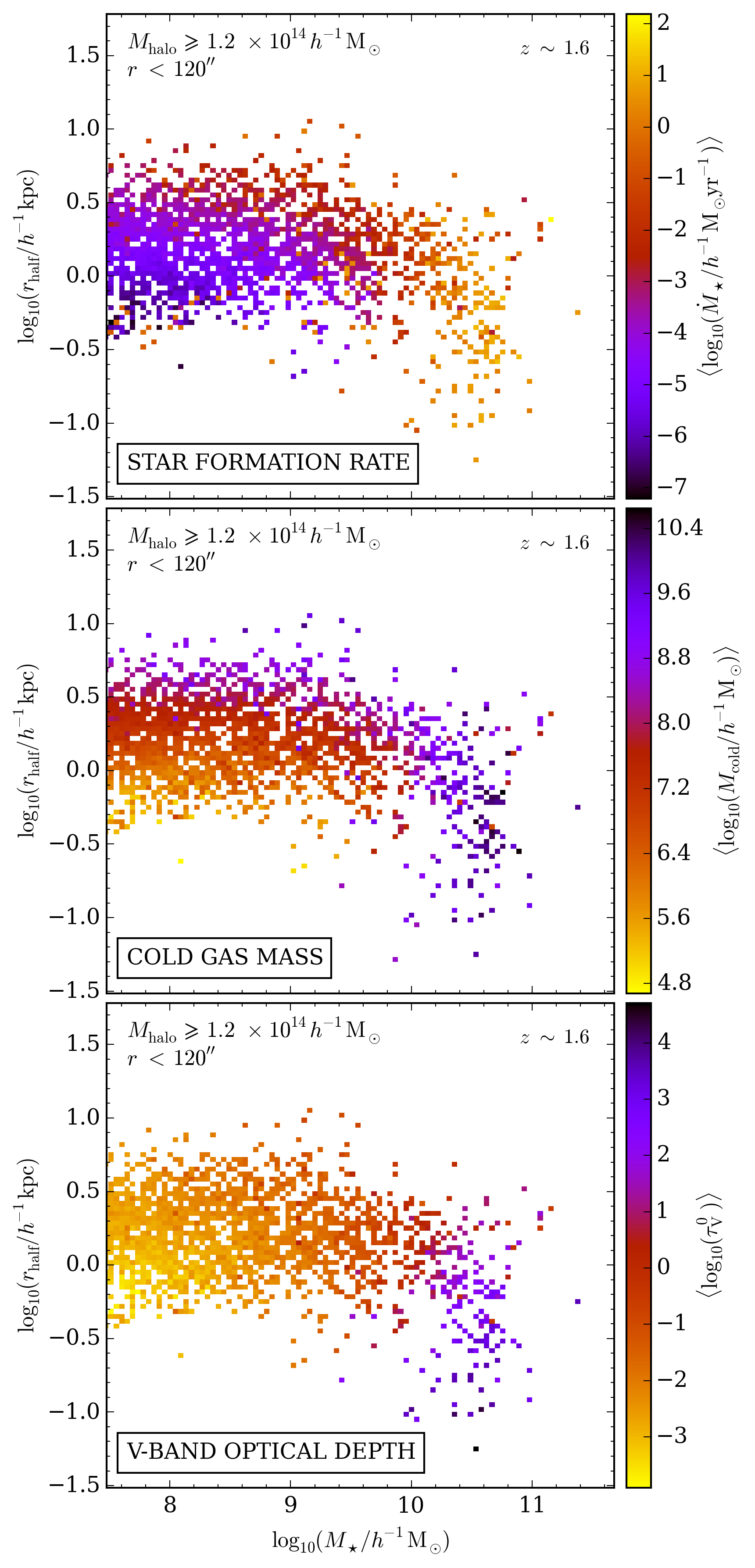}
  \caption{Distribution of selected galaxy properties along the
    mass-size relation as predicted by \GALFORM{}. The various
    properties: star formation rate, half-mass radius and
    \band{V}-band optical depth, are labelled in the corresponding
    panel. Note that the half-mass radius includes both the disc and
    bulge of the galaxy and the optical depth is the depth measured
    through the centre of the galaxy when face-on. The pixels are
    coloured according to the median value of the galaxy property for
    the galaxies in that pixel. The redshift, as well as the halo mass
    and aperture used to select the \GALFORM{} galaxies, are shown in
    each panel.}
  \label{fig:size_mass_props}
\end{figure}

Perhaps the easiest of these properties to compare against
observations is the size of the galaxies. In Fig.~\ref{fig:size_mass}
we compare the stellar mass-size relation for the model cluster
galaxies with observational estimates from \citet{Szomoru12} at
$z\sim1.2$ and from \citet{Papovich12} and \citet{Saracco14} at
$z\sim1.6$. Note that the galaxies we show from
\citeauthor{Papovich12} are those selected as cluster galaxies,
whereas the galaxies from \citeauthor{Szomoru12} and
\citeauthor{Saracco14} are from global samples\footnote{To convert the
  stellar mass estimates from \citeauthor{Papovich12} and
  \citeauthor{Saracco14} from the \citeauthor{Chabrier03} initial mass
  function (IMF) to the \citet{Kennicutt83} IMF we adopt a conversion
  factor of -0.09 dex \citep{Mitchell13}. To convert the stellar
  masses estimates of \citeauthor{Szomoru12}, we first converted from
  the \citet{Kroupa01} IMF to the \citet{Salpeter55} IMF using a
  conversion factor of -1.6 dex \citep{Fontana04} and then converted
  from the \citet{Salpeter55} IMF to the \citet{Kennicutt83} IMF using
  a conversion factor of +1.4 dex \citep{Fontana04}.}. For stellar
masses $M_{\star}\lesssim10^{10}h^{-1}{\rm M_{\odot}}$, the mass-size
relation for the model, which is shown by the greyscale pixels, is
approximately flat but with a large scatter of typically $0.4-0.5$
dex. The mass-size relation drops off towards the highest
masses. Therefore, although there is agreement between the
observations and the model predictions, the most massive cluster
galaxies, with stellar masses $M_{\star}\gtrsim10^{10}h^{-1}{\rm
  M_{\odot}}$, are typically smaller than the observed sizes, with
some model galaxies being up to an order of magnitude
smaller. Examining how the stellar mass-size relation correlates with
other galaxy properties, we can see in Fig.~\ref{fig:size_mass_props}
that the population of very massive, compact galaxies in the model
typically have the highest optical depth, which is not surprising
given Eq.~\ref{eqn:optical_depth}. We also find that these high-mass,
compact galaxies have extremely large cold gas masses of $M_{{\rm
    cold,gas}}\sim10^{10}h^{-1}{\rm M_{\odot}}$ (comparable to their
stellar masses) as well as the highest star formation rates (between
$1h^{-1}{\rm M_{\odot}}{\rm yr}^{-1}$ and $10h^{-1}{\rm M_{\odot}}{\rm
  yr}^{-1}$).

Overall, the large reservoirs of cold gas and the compact sizes of the
most massive cluster galaxies appear to be the cause of the large dust
attenuation, which leads to the discrepancy between the observed CGLF
and that predicted by the model. Our results therefore suggest that
the problem of the large dust attenuation is actually due to an
underlying problem of either the model under-predicting the sizes of
the most massive cluster galaxies or the model allowing too much
cooling of gas in these halos. \citet{Gonzalez09} come to a similar
conclusion when comparing predictions from the \citet{Baugh05} and
\citet{Bower06} \GALFORM{} models with the colours of galaxies in the
Sloan Digital Sky Survey
\citep[SDSS,][]{York00}. \citeauthor{Gonzalez09} find that the model
over-predicts the number of bright, blue galaxies and also that the
bulge-dominated bright galaxies have sizes up to a factor of ten times
smaller than SDSS galaxies of equivalent luminosity. Following an
examination of several of the model parameters, they attribute the
problem as being due to an over-simplified treatment of the sizes of
galaxy merger remnants. At lower redshift, \citet{Weinzirl14} compare
the predictions of the semi-analytical model of \citet{Neistein10}
with Hubble Space Telescope observations of the Coma Cluster and find
that the model over-predicts the mass fraction of cold gas for
galaxies in halos with Coma-like properties.

\subsection{Robustness of model predictions to parameter changes}
\label{sec:galform_parameters}

In the previous section we concluded that the discrepancy between the
observed CGLF and the model predictions is likely being caused by the
model predicting too much cold gas and too small sizes for the most
massive cluster galaxies. Since these two fundamental galaxy
properties are sensitive to many other model parameters, we now vary
some of the key parameters that we would expect to have the greatest
effect upon the high redshift cluster galaxy population in order to
gain some insight into the physics shaping the model predictions for
cluster galaxies. We note that most of the parameters we choose to
vary will have a greater impact on the cold gas masses of the galaxies
rather than the galaxy sizes. For this exercise we will vary each
parameter independently and use the prediction for the $z\sim1.4$ CGLF
as an indicator of possible improvements, since the discrepancy is
most noticeable in the predictions for the CGLF. We note, however,
that for more a extensive search varying multiple parameters
simultaneously and examination of multiple galaxy statistics would be
necessary. We leave such a search for future work.

The main parameters that we expect to have the greatest influence on
the predictions for the cluster galaxy population are those governing
the heating and cooling of gas in the most massive halos as well as
those parameters governing treatment of galaxy mergers. Besides these
parameters, we also examined varying the scale height of the dust in
the model galaxies. Although this parameter cannot improve the model
predictions for the galaxy sizes or cold gas masses, it may allow an
improved recovery of the CGLF. We find, however, that this has limited
impact upon the model prediction for the CGLF.

\subsubsection{Galaxy mergers and interactions}
\label{sec:galaxy_mergers}

In \GALFORM{} spheroids are created following galaxy mergers and disc
instabilities. These events can also trigger starburst events, which
would act to deplete the cold gas reservoirs of the merger remnant. We
might expect therefore that varying the parameter governing the
timescale of these starbursts would have an effect upon the CGLF. We
find, however, that changing the duration of starbursts has little
impact upon the predicted CGLF.

From their analysis, \citet{Gonzalez09} found that changing the
prescription for calculating the size of the stellar spheroid
following a galaxy merger had a large impact upon the sizes of bright
elliptical galaxies at low redshift. However, when we adopt their suggested
parameter values we see little change in the CGLF or the predicted
sizes for the most massive cluster galaxies. This might suggest that
the amount of stellar mass produced in galaxy mergers is less
important in high redshift cluster galaxies compared to the local
Universe, which agrees with the lack of sensitivity we have seen to
the starburst timescales.

\citet{Font08} demonstrated that the incorporation of gradual
ram-pressure stripping into \GALFORM{} improves the model predictions
for the colours of satellite galaxies compared with
observations. Recently, \citet{Lagos14} have also shown that gradual
ram pressure stripping is needed to reproduce the atomic and molecular
gas contents of early-type galaxies. In \GALFORM{}, when galaxies
become satellites they have their hot gas stripped
instantaneously. \citet{Font08} included a prescription to delay this
stripping and allowed satellite galaxies to retain a fraction of their
hot gas for a longer period, thus delaying the quenching of their star
formation. However, when we include the \citeauthor{Font08} treatment
for stripping, we again see a negligible change in the prediction for
the CGLF, which might suggest that the galaxy sizes are having the
greatest impact upon the dust attenuation calculation.

In our reference model, the merger timescale for a satellite galaxy is
calculated, based upon dynamical friction arguments, every time its
host halo undergoes a merger event. The satellite is assumed to merge
onto the central galaxy after this time, irrespective of whether the
sub-halo hosting the galaxy is still identifiable in the
simulation. \citet{Campbell14} recently showed that an alternative
scheme, where the merger timescale is computed once the host sub-halo
can no longer be identified, leads to a change in the model
predictions for the stellar mass function at $z=0$. We find that
adopting this scheme does not improve the model predictions for the
CGLF.

\begin{figure}
  \centering
  \includegraphics[width=0.46\textwidth]{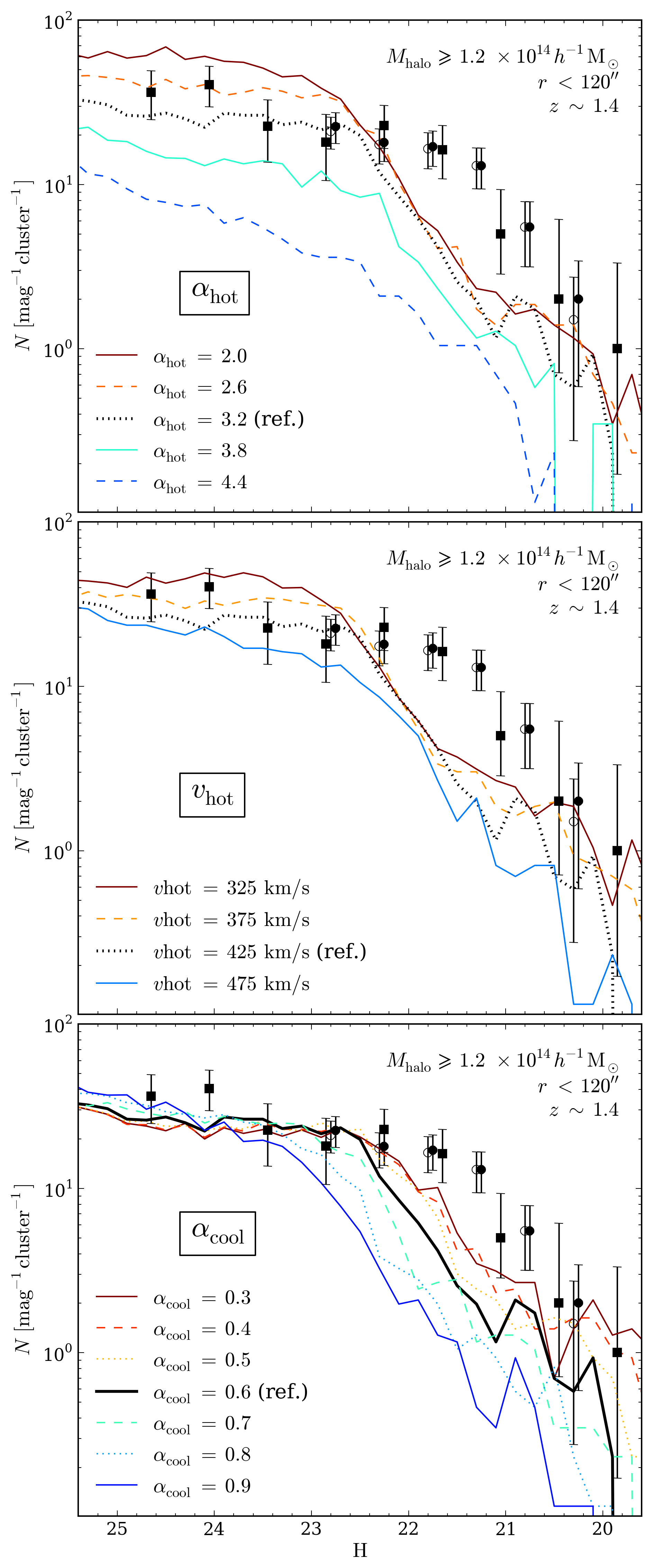}
  \caption{Impact on \band{H}-band cluster galaxy luminosity function
    at $z\sim1.4$ as predicted by the \GALFORM{} model when varying
    the parameters governing SN and AGN feedback. The top panel shows
    the predictions when varying the SN feedback parameter
    $\alpha_{{\rm hot}}$, the middle panel shows the predictions when
    varying the SN feedback parameter $v_{{\rm hot}}$ and the bottom
    panel shows the predictions when varying the AGN feedback
    parameter $\alpha_{{\rm cool}}$. The parameters used by our
    reference model are indicated in the legend of each panel. All
    predictions assume the dust attenuation calculation described in
    $\S$~\ref{sec:GALFORM_dust}. The halo mass and aperture used to
    select the semi-analytical cluster galaxies are indicated in each
    panel. The data points are the same as from the right-hand panel
    of Fig.~\ref{fig:cglf_hband}.}
  \label{fig:HCGLF_feedback}
\end{figure}

\subsubsection{Supernovae feedback}
\label{sec:varying_supernovae}

The amount of cold gas in massive cluster galaxies could be lowered by
reducing the strength of feedback due to supernovae such that the
galaxies undergo more star formation at earlier epochs, prior to them
falling into the clusters. However, feedback due to supernovae is
thought to affect the faint-end slope of the global galaxy luminosity
function \citep[e.g.][]{Benson03}. As such, it is possible that
reducing the strength of the supernova feedback will reduce the gas
content of the galaxies, but will produce an undesirable boost in the
faint-end of the CGLF above the observations.

In the upper two panels of Fig.~\ref{fig:HCGLF_feedback} we show the
effect of independently varying the free parameters
$\alpha_{{\rm hot}}$ and $v_{{\rm hot}}$. Variation of either
parameter clearly affects the normalisation of the
CGLF. Understandably, the change in normalisation is greater at the
faint-end. For the bright-end of the CGLF, there is little change in
the normalisation as the parameters are decreased below their fiducial
values, suggesting that further change in the supernova feedback alone
would have minimal impact. The values for the parameters could be
reduced further beyond the range considered in
Fig.~\ref{fig:HCGLF_feedback}, but this would cause an excess in the
faint-end above the observations. As such, we conclude that changing
the strength of the supernova feedback alone is unable to fix the
deficit around the break in the CGLF.

\begin{figure*}
  \centering
  \includegraphics[width=0.96\textwidth]{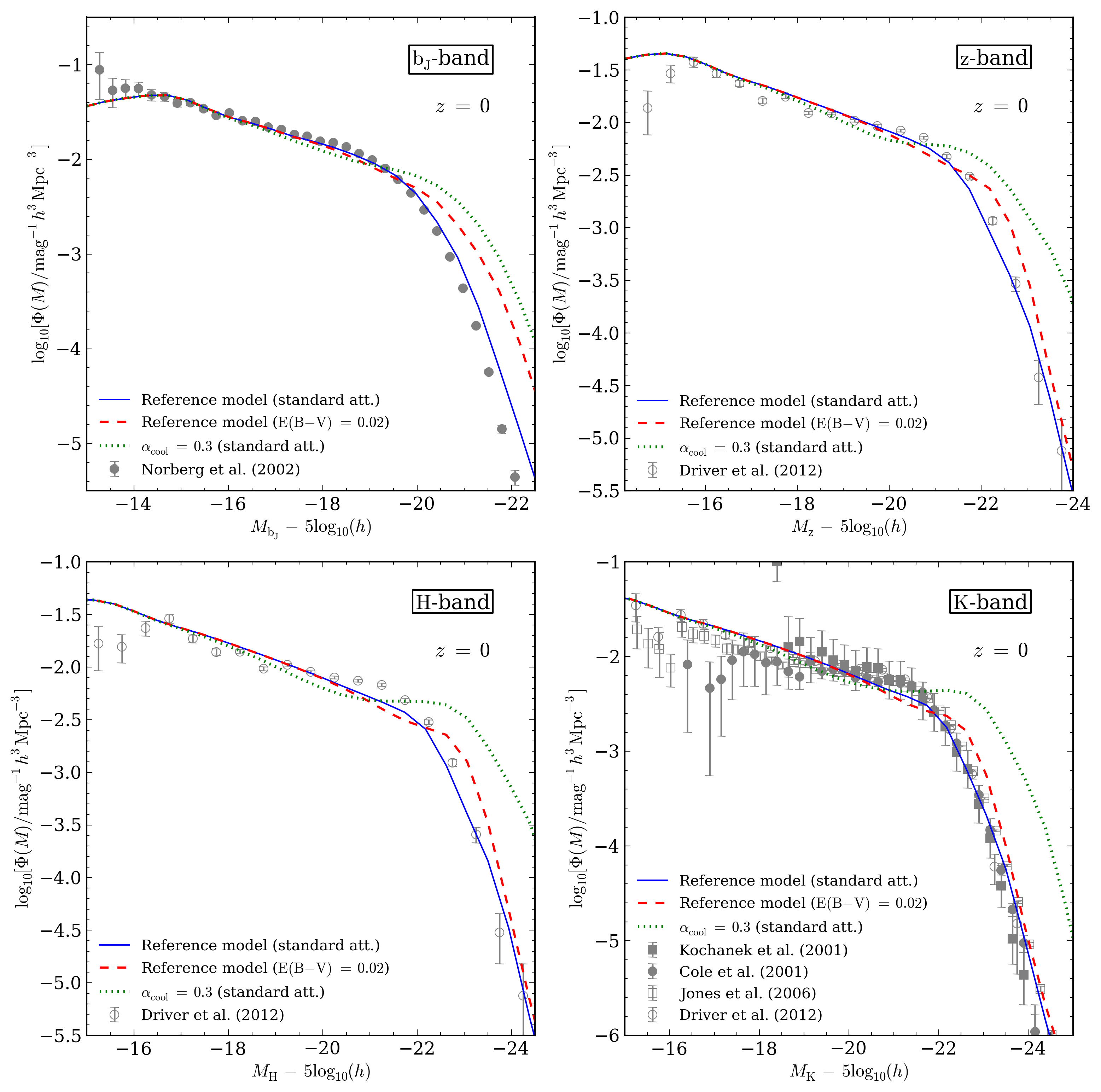}
  \caption{Predicted luminosity functions at $z=0$ for the
    \emph{global} galaxy population (both field and cluster
    galaxies). The panels show the luminosity functions in the ${\rm
      b_J}$, ${\rm z}$, ${\rm H}$ and ${\rm K}$-bands, as
    labelled. Data points show the observational estimates from the
    Two-Micron All-Sky Survey \protect\citep[2MASS,][]{Kochanek01},
    Two-degree Field Galaxy Redshift Survey 2dFGRS
    \protect\citep[2dFGRS,][]{Cole01,Norberg02}, the Six-degree Field
    Galaxy Survey \protect\citep[6dFGS,][]{Jones06} and the Galaxy And
    Mass Assembly Survey \protect\citep[GAMA,][]{Driver12}. The solid
    blue line shows the prediction for the fiducial \GALFORM{} model
    \protect\citep{Gonzalez-Perez14} with the default dust attenuation
    calculations. The dashed line shows the prediction of the fiducial
    model when, instead of the default extinction, a
    \protect\citet{Calzetti00} law with ${\rm E(B-V)}=0.02$ is
    adopted. The green, dotted line shows the \GALFORM{} prediction,
    with the default dust extinction, when the parameter $\alpha_{{\rm
        cool}}$ is reduced from $0.6$ to $0.3$.}
  \label{fig:global_z0_lf}
\end{figure*}

\subsubsection{AGN feedback}
\label{sec:varying_agn}

Feedback due to active galactic nuclei (AGN) is expected to have a
dramatic impact on galaxies residing in relatively massive systems,
like galaxy clusters, through the quenching of any active
star formation.

\citet{Bower06} and \citet{Croton06} were amongst the first to
demonstrate that introducing feedback due to AGN into semi-analytic
models helps reduce the number of bright galaxies in the models, thus
improving the match to the bright-end of the observed global galaxy
luminosity function. Additionally the action of AGN feedback leads to
the models predicting a bi-modal colour-magnitude relation, similar to
that observed in the SDSS \citep{Gonzalez09}. Since the bright end of
the global luminosity function is dominated by cluster galaxies, we
expect that adjusting the parameters controlling AGN feedback will
affect the predicted CGLF.

As discussed in $\S$\ref{sec:GALFORM}, in \GALFORM{} the strength of
AGN feedback is governed by the parameter $\alpha_{{\rm cool}}$. The
effect of varying the value of $\alpha_{{\rm cool}}$ on the cluster
galaxy luminosity function is shown in the lower panel of
Fig.~\ref{fig:HCGLF_feedback} where we plot the \band{H}-band CGLF for
a range of values of $\alpha_{{\rm cool}}$, either side of the value
of $\alpha_{{\rm cool}}=0.6$ used in the reference model. As expected,
we see that variations in the value of $\alpha_{{\rm cool}}$ has a
dramatic effect upon the sharpness of the break and normalisation of
the bright-end of the CGLF. In contrast, the change in the
normalisation of the faint-end of the CGLF is negligible, especially
when $\alpha_{{\rm cool}}$ is reduced below the reference value.

We can see from Fig.~\ref{fig:HCGLF_feedback} that reducing
$\alpha_{{\rm cool}}$ brings the model predictions into better
agreement with the observations, though even for $\alpha_{{\rm
    cool}}\lesssim0.4$ there is still a slight discrepancy around the
break in the CGLF, suggesting that a further reduction in
$\alpha_{{\rm cool}}$ would be necessary. Examination of the CMR shows
that adopting $\alpha_{{\rm cool}}\lesssim0.4$ has negligible impact
upon the galaxy colours, with the slope and the zero-point of the RS
(fitted to those galaxies with $\log_{10}({\rm
  SFR}/h^{-1}\Msolyr)\leqslant-2$) changing on the order of one per
cent. However, as we shall see in $\S$\ref{sec:z0_predictions},
assuming such weak AGN feedback has a significant impact upon the
model predictions at $z=0$, which indicates that solely reducing AGN
feedback is not an adequate solution and that some other mechanism
must be changed, or introduced, if the model is to correctly predict
the colours and abundances of high redshift cluster galaxies.

\subsubsection{Influence on local Universe predictions}
\label{sec:z0_predictions}

Following our brief parameter search we have found that the strength
of AGN feedback could be used to reduce the discrepancy between the
model predictions and observations of high redshift galaxy
clusters. We now examine how our attempt to match observations of high
redshift clusters changes the predictions of the model at $z=0$, which
were originally used to calibrate the model.

One of the principal statistics used to constrain the parameters of
the \GALFORM{} model is the \emph{global} galaxy luminosity function
(of both field and cluster galaxies) at $z=0$, specifically in the
$\band{b_J}$ and $\band{K}$-bands. In Fig.~\ref{fig:global_z0_lf} we
show the global galaxy luminosity function at $z=0$ for the
$\band{b_J}$, $\band{z}$, $\band{H}$ and $\band{K}$ bands. The
predictions for our reference \GALFORM{} model (adopting the default
dust attenuation calculation) are shown by the solid line. Since the
parameters of the fiducial model have been constrained using the
$\band{b_J}$ and \band{K} band luminosity functions, this model
provides a good match to the global luminosity function in each of the
four bands.

The red, dashed line in Fig.~\ref{fig:global_z0_lf} shows the impact
at $z=0$ of assuming weaker dust attenuation, in this case a
\citet{Calzetti00} law with ${\rm E(B-V)}=0.02$, which is gives a
better match to the CGLF. As expected, weaker dust attenuation boosts
the abundance of bright galaxies, with a reduction in the predicted
counts around the knee of the luminosity function. The impact on the
luminosity function of adopting weaker dust attenuation becomes more
significant as one moves from the near-infrared towards the optical,
with the \band{K}-band showing the smallest change and the
$\band{b_J}$-band showing the largest. We have seen in
$\S$\ref{sec:varying_agn} that weaker AGN feedback is necessary to
reconcile the model predictions for the CGLF and the observations. As
such, we also plot in Fig.~\ref{fig:global_z0_lf} the predicted $z=0$
luminosity function for the model when $\alpha_{{\rm cool}}=0.3$ is
assumed. The effect on the luminosity function is dramatic, with the
weaker AGN feedback leading to a significant excess of bright
galaxies.

These latter two predictions again highlight the challenge facing
current galaxy formation models. In addition to reproducing the sizes,
luminosities and colours of massive galaxies in clusters, the models
need to be able to remain consistent with observations of the local
Universe. To achieve this requires more than one parameter to be
varied as well as the possible inclusion of new physics.

%%%%%%%%%%%%%%%%%%%%%%%%%%%%%%%%%%%%%%%%%%%%%%%%%%%%%%%%%%%%%%%%%%%%%%%%%%%%%%%%%%%%%%%%%%%%%%%%%%%
%%%%%%%%%%%%%%%%%%%%%%%%%%%%%%%%%%%%%%%%%%%%%%%%%%%%%%%%%%%%%%%%%%%%%%%%%%%%%%%%%%%%%%%%%%%%%%%%%%%

\section{Summary \& conclusions}
\label{sec:conclusions}

We have compiled observations of high redshift ($z>1$) galaxy
clusters, which we compare to the predictions of the
\citet{Gonzalez-Perez14} variant of the \GALFORM{} semi-analytical
galaxy formation model, which we treat as our reference model. The
statistics that we consider are the cluster galaxy luminosity function
(CGLF) and the colour-magnitude relation (CMR). To identify cluster
galaxies in the semi-analytic catalogue, we select only those galaxies
in halos with mass greater than $1.2\times 10^{14}h^{-1}\Msol$. We
further use the distant observer approximation to apply an aperture
and reject those galaxies lying further than $120^{\arcseconds}$ away
from the halo centre.

Our reference \GALFORM{} model predicts a CGLF in reasonable agreement
with the observed CGLF at the faint and bright ends, but significantly
under-predicts the number of cluster galaxies around the break in the
CGLF. Examination of several possible factors that might affect the
model predictions, including aperture size and halo mass selection,
indicates that the discrepancy between the observations and the model
predictions is likely caused by the reference model applying an overly
large dust attenuation. If we instead apply a weaker dust attenuation,
which we represent using a \citet{Calzetti00} law, then the reference
model prediction is able to provide a much better fit to the observed
CGLF. We note that we do not advocate that a \citeauthor{Calzetti00}
law is the correct description for dust attenuation at high redshift,
but instead have simply used the law to demonstrate the impact of the
large dust attenuation predicted in our reference model.

In contrast, the reference \GALFORM{} model predicts a CMR that is
qualitatively consistent with the observed colours of cluster galaxies
at $z\sim1.2$, $z\sim1.4$ and $z\sim1.6$. We provide linear fits to
the red sequence, RS, for different colour-spaces using a subset of
passively evolving galaxies selected using the sSFR cut,
$\log_{10}(\mathrm{sSFR}/\mathrm{Gyr}^{-1})\leqslant-1$. The slopes
and zero-points of these fits are broadly consistent with
observationally derived estimates.

The CMR predicted by our reference model displays a subset of very red
galaxies, which appear in a `plume' above the predicted RS. We
determine that these galaxies are a result of the large dust
attenuation that is causing the discrepancy between the observed CGLF
and that in the model. However, although a weaker dust attenuation
improves the model predictions for the CGLF, assuming a weaker dust
attenuation worsens the predicted RS, with the creation of a branch of
blue galaxies extending below the RS.

Examination of the properties of those galaxies with large dust
attenuation reveals them to be highly star-forming, with large amounts
of cold gas and with small scale sizes. The large reservoirs of cold
gas and the compact sizes appear to be the cause of the large
predicted dust attenuation. To gain some insight into this problem we
briefly examined how varying several key model parameters changes the
predicted cluster statistics, in particular the CGLF. We find that a
reduction in the strength of feedback due to AGN is able to provide
some improvement in the form of the CGLF, but makes the colours of the
model galaxies too blue. In addition, introducing a weaker AGN
feedback significantly affects the model predictions at $z=0$ by
boosting the number of bright galaxies and over-predicting the counts
at the bright end of $z=0$ global galaxy luminosity function.

There are several possible explanations for the discrepancy between
the observations and the model predictions. Amongst the most probable
are that \GALFORM{} is under-predicting the star formation of these
massive cluster galaxies, or allowing too much gas cooling, which
would leave them too faint (and most likely too blue) and with lots of
cold gas by the time they become cluster galaxies. The problem of
semi-analytical models under-predicting star formation at high
redshift has been commented on several times in the literature
\citep[e.g.][]{Daddi07,Damen09,Fontanot09b,Dutton10,Weinmann11,
  Weinmann12}. Similar deficiencies have also been reported in
hydrodynamical simulations \citep[e.g.][]{Kannan14,Wang15}. However,
it is also worth noting a recent result from \citet{Chang15} who
suggest that observational calibration of SFR estimates could be wrong
by approximately a factor of two, which would result in previous
observational SFR estimates being a factor of two too large. If this
is the case then correcting for this would bring the observational SFR
estimates and theoretical predictions into closer
agreement. Additionally, the under-prediction of the sizes of the
galaxies in \GALFORM{} is likely to be having a significant impact
upon the other model predictions. Predicting correct galaxy sizes is a
long standing problem for semi-analytical models since it is entangled
with the contraction of the host halo. These conclusions are in
agreement with previous comparisons to semi-analytical model
predictions \citep[e.g.][]{Gonzalez09, Weinzirl14}.

Overall, these results demonstrate the challenge facing current galaxy
formation models such as \GALFORM{}: how to match the luminosity
abundance and colours of cluster galaxies, whilst remaining consistent
with the observed properties of galaxies in the local
Universe. Achieving this will require incorporating into the models a
better understanding of galaxy evolution in extreme
environments. However, we must keep in mind that the comparisons in
this work are limited to a small sample of individual galaxy clusters
and that understanding of the astrophysical processes affecting the
high redshift galaxy population remains uncertain. Ultimately, our
understanding of the abundance and properties of high redshift
clusters will only improve as we improve our statistics with up and
coming deep, wide-field galaxy surveys such as DES and Euclid.

%%%%%%%%%%%%%%%%%%%%%%%%%%%%%%%%%%%%%%%%%%%%%%%%%%%%%%%%%%%%%%%%%%%%%%%%%%%%%%%%%%%%%%%%%%%%%%%%%%%
%%%%%%%%%%%%%%%%%%%%%%%%%%%%%%%%%%%%%%%%%%%%%%%%%%%%%%%%%%%%%%%%%%%%%%%%%%%%%%%%%%%%%%%%%%%%%%%%%%%

\section*{Acknowledgements}
We thank the referee for a thorough report with many useful and
constructive comments. Thanks go to Rene Fassbender, Chris Lidman,
Adam Stanford, Veronica Strazzullo and Gregory Snyder for provision of
their data. In addition, we thank Bego\~{n}a Ascaso, James Bartlett,
Andrea Biviano, Stefano Borgani, Mark Brodwin, Alberto Cappi, Anthony
Gonzalez, Sophie Maurogordato, Pierluigi Monaco, David Murphy and Adam
Stanford for many useful and productive discussions and
suggestions. This work was motivated by feedback from members of the
Euclid Consortium Galaxy Clusters Science Working Group following the
provision of galaxy mock catalogues based upon the \GALFORM{}
model. VGP acknowledges support from a European Research Council
Starting Grant (DEGAS-259586). FBA acknowledges the support of the
Royal Society for a University Research Fellowship. This work carried
out on the COSMA Data Centric system at Durham University, operated by
the Institute for Computational Cosmology on behalf of the STFC DiRAC
HPC Facility ([http://]www.dirac.ac.uk). This equipment was funded by
a BIS National E-infrastructure capital grant ST/K00042X/1, DiRAC
Operations grant ST/K003267/1 and Durham University. DiRAC is part of
the National E-Infrastructure.

%%%%%%%%%%%%%%%%%%%%%%%%%%%%%%%%%%%%%%%%%%%%%%%%%%%%%%%%%%%%%%%%%%%%%%%%%%%%%%%%%%%%%%%%%%%%%%%%%%%
%%%%%%%%%%%%%%%%%%%%%%%%%%%%%%%%%%%%%%%%%%%%%%%%%%%%%%%%%%%%%%%%%%%%%%%%%%%%%%%%%%%%%%%%%%%%%%%%%%%

\bibliographystyle{mn2e_trunc8}
\bibliography{aimerson}

%%%%%%%%%%%%%%%%%%%%%%%%%%%%%%%%%%%%%%%%%%%%%%%%%%%%%%%%%%%%%%%%%%%%%%%%%%%%%%%%%%%%%%%%%%%%%%%%%%%
%%%%%%%%%%%%%%%%%%%%%%%%%%%%%%%%%%%%%%%%%%%%%%%%%%%%%%%%%%%%%%%%%%%%%%%%%%%%%%%%%%%%%%%%%%%%%%%%%%%

\end{document}

%% file: rsfits_table_accepted.tex
\begin{table}
\centering
\caption{Fits to the red sequences predicted by the
  \citet{Gonzalez-Perez14} \GALFORM{} model for various
  colour-magnitude spaces between redshifts $z=1.2$ and $z=1.6$ using
  subsets of model cluster galaxies selected by a cut in
  specific star formation rate, sSFR (a passively evolving sample). Note that the
  samples were additionally limited to galaxies
  brighter than $25^{\mathrm{th}}$ magnitude.}
\begin{tabular}{| c | c | c | c |}
\hline
Magnitude&Colour&Slope&Zero-point\\
\hline\hline
\multicolumn{4}{|l|}{REDSHIFT: 1.2}\\
$\band{z}$&$\band{i}-\band{z}$&$-0.0118\pm0.0005$&$1.0244\pm0.0009$\\
$\band{H}$&$\band{i}-\band{H}$&$-0.146\pm0.002$&$2.436\pm0.003$\\
$\band{K}$&$\band{J}-\band{K}$&$-0.118\pm0.001$&$0.740\pm0.002$\\
\hline
\multicolumn{4}{|l|}{REDSHIFT: 1.4}\\
$\band{J}$&$\band{z}-\band{J}$&$-0.060\pm0.003$&$1.348\pm0.004$\\
$\band{H}$&$\band{z}-\band{H}$&$-0.115\pm0.004$&$1.839\pm0.005$\\
$\band{K}$&$\band{J}-\band{K}$&$-0.112\pm0.003$&$0.849\pm0.004$\\
\hline
\multicolumn{4}{|l|}{REDSHIFT: 1.6}\\
$\band{K}$&$\band{J}-\band{K}$&$-0.106\pm0.007$&$0.99\pm0.01$\\
\hline
\end{tabular}
\label{tab:RS_fits}
\end{table}